\documentclass[journal,10pt]{IEEEtran}
\usepackage{graphicx, enumerate, amsmath, yhmath, amssymb, psfrag,algorithm, algorithmic,xspace,tikz,cancel}
\usepackage{soul}
\usepackage[caption=false,font=footnotesize]{subfig}
\usetikzlibrary{shapes,arrows,dsp,chains}
\tikzstyle{line} = [draw, -latex']

\newcommand{\sipp}{\ensuremath{\textsc{sipp}}\xspace}
\newcommand{\dipp}{\ensuremath{\textsc{dipp}}\xspace}

\newcommand{\cs}{\textsc{cs}\xspace}
\newcommand{\dcs}{\textsc{dcs}\xspace}
\newcommand{\expansion}{\ensuremath{\texttt{expansion}}\xspace}
\newcommand{\consensus}{\ensuremath{\texttt{consensus}}\xspace}
\renewcommand{\sp}{\ensuremath{\textsc{sp}}\xspace}
\newcommand{\srer}{\ensuremath{\textsc{srer}}\xspace}
\newcommand{\supp}{\ensuremath{\texttt{supp}}}
\newcommand{\mbx}{\mathbf{x}}
\newcommand{\mbA}{\mathbf{A}}
\newcommand{\mby}{\mathbf{y}}
\newcommand{\mbe}{\mathbf{e}}
\newcommand{\mbr}{\mathbf{r}}
\newcommand{\mbq}{\mathbf{q}}

\renewcommand{\complement}{\ensuremath{\mathsf{c}}}
\newenvironment{packed_enum}{
\begin{enumerate}
}{\end{enumerate}}
\newtheorem{definition}{Definition}
\newtheorem{example}{Example}

\newtheorem{proposition}{Proposition}
\newtheorem{remark}{Remark}
\newtheorem{corollary}{Corollary}
\newtheorem{lemma}{Lemma}

\newtheorem{assumption}{Assumption}

\newcommand{\qed}{\hfill \ensuremath{\blacksquare}}
\newcommand{\wqed}{\hfill \ensuremath{\square}}
\newcommand{\addi}{\texttt{vote}_1}
\newcommand{\resid}{\texttt{res}}

\newcommand{\T}{\mathcal{T}}
\newcommand{\Th}{\hat{\mathcal{T}}}
\renewcommand{\S}{\mathcal{S}}

\newcommand{\I}{\mathcal{I}}
\newcommand{\p}{\ensuremath{\mathsf{p}}}
\newcommand{\q}{\ensuremath{\mathsf{q}}}
\renewcommand{\r}{\ensuremath{\mathsf{r}}}
\newcommand{\Ti}{\ensuremath{\mathcal{T}_{\p}}}

\newcommand{\Thi}{\ensuremath{\hat{\mathcal{T}}_{\p}}}
\newcommand{\Thii}{\ensuremath{\hat{\mathcal{T}}_{\q}}}
\newcommand{\Thiii}{\ensuremath{\hat{\mathcal{T}}_\r}}
\newcommand{\J}{\ensuremath{\mathcal{J}}}
\newcommand{\Jh}{\ensuremath{\hat{\mathcal{J}}}}
\newcommand{\Ii}{\ensuremath{\mathcal{I}_{\p}}}

\newcommand{\asce}{\textsc{asce}\xspace}
\newcommand{\gp}{\textsc{gp}\xspace}
\newcommand{\C}{\mathbf{C}}
\newcommand{\omp}{\textsc{omp}\xspace}
\newcommand{\cosamp}{\textsc{c}o\textsc{s}a\textsc{mp}\xspace}
\newcommand{\smnr}{\textsc{smnr}\xspace}

\renewcommand{\appendixname}{appendix\xspace}
\newcommand{\sectionname}{section}
\newcommand{\algorithmname}{algorithm}
\newcommand{\propositionname}{proposition}
\newcommand{\lemmaname}{lemma}
\newcommand{\corollaryname}{corollary}
\newcommand{\assumptionname}{assumption}
\renewcommand{\figurename}{Fig.}

\renewcommand{\P}[1]{\ensuremath{\mathbb{P}\,\!\bigl(#1\bigr)}}

\makeatletter
\newcommand{\gettikzxy}[3]{%
  \tikz@scan@one@point\pgfutil@firstofone#1\relax
  \edef#2{\the\pgf@x}%
  \edef#3{\the\pgf@y}%
}
\makeatother

\begin{document}
%
\title{
{{Design and Analysis of a Greedy Pursuit \\ for Distributed Compressed Sensing}} \\
}
%
%
%

\author{Dennis~Sundman*,~\IEEEmembership{Student~Member,~IEEE,}
        Saikat~Chatterjee,~\IEEEmembership{Member,~IEEE,}
        and~Mikael~Skoglund,~\IEEEmembership{Senior~Member,~IEEE}
\thanks{The authors are with the Communication Theory Department, KTH Royal Institute of Technology, 10044 Stockholm, Sweden e-mail: (denniss@kth.se, sach@kth.se, and skoglund@kth.se).}
\thanks{Manuscript received April 19, 2005; revised January 11, 2007.}%
\thanks{This paper was presented in part at 1st IEEE Global Conference on Signal and Information Processing 2013~\cite{Sundman:parallel_pursuit_for_dcs}.}}

%
%

\markboth{Journal of \LaTeX\ Class Files,~Vol.~6, No.~1, January~2007}%
{Shell \MakeLowercase{\textit{et al.}}: Bare Demo of IEEEtran.cls for Journals}
%



\maketitle

\begin{abstract}

We consider a distributed compressed sensing scenario where many sensors measure correlated sparse signals and the sensors are connected through a network. {{Correlation between sparse signals is modeled by a partial common support-set. For such a scenario, the main objective of this paper is to develop a greedy pursuit algorithm. We develop a distributed parallel pursuit (\dipp) algorithm based on exchange of information about estimated support-sets at sensors. The exchange of information helps to improve estimation of the partial common support-set, that in turn helps to gradually improve estimation of support-sets in all sensors, leading to a better quality reconstruction performance. We provide restricted isometry property (RIP) based theoretical analysis on the algorithm's convergence and reconstruction performance. 
Under certain theoretical requirements on the quality of information exchange over network and RIP parameters of sensor nodes, we show that the \dipp algorithm converges to a performance level that depends on a scaled additive measurement noise power (convergence in theory) where the scaling coefficient is a function of RIP parameters and information processing quality parameters. Using simulations, we show practical reconstruction performance of \dipp vis-a-vis amount of undersampling, signal-to-measurement-noise ratios and network-connectivity conditions.}} 

\end{abstract}


%
\IEEEpeerreviewmaketitle

\section{Introduction}
%
%
%
%

\IEEEPARstart{C}{}{ompressed} sensing (\cs)~\cite{Donoho:compressed_sensing,Candes:stable_signal_recovery} refers to a class of under-sampling problems, where the sampled (or measured) data is inherently sparse. 
A standard \cs problem typically considers a single-sensor scenario, where the main task is reconstruction of a large-dimensional signal-vector from a small-dimensional measurement-vector by using a-priori knowledge that the signal is sparse in a known domain. 
Several \cs reconstruction algorithms have been developed in the literature, for example convex optimization- \cite{Mota:distributed_basis_pursuit, Bazerque:distributed_spectrum_sensing} and Bayesian- \cite{Ji:bayesian_compressive_sensing, Donoho:message_passing_for_cs} algorithms. 
An important class of reconstruction algorithms is the greedy pursuits (\gp)
which are popular to use for large \cs problems due to a good trade-off between computational complexity and reconstruction performance. 
From a measurement vector, the \gp algorithms use simple linear algebraic tools to estimate the underlying support-set of the sparse signal-vector followed by estimating associated signal values on the support-set. Here we mention that a good support-set estimation is an important engineering aspect for the \gp algorithms. 
Considering support-set estimation strategy, \gp algorithms can be categorized in two broad classes: sequential and parallel. Sequential strategy estimates a support-set by finding elements of the support-set one-by-one over iterations. On the other hand, parallel strategy estimates all elements of a support-set simultaneously in an iteration, but improves the support-set estimate over iterations. For example, the sequential types include matching pursuit \cite{Mallat:matching_pursuit_with_time_frequency_dictionaries}, orthogonal matching pursuit (\textsc{omp})~\cite{Tropp:signal_recovery}, and their algorithmic variations \cite{Donoho:sparse_solution_of_underdetermined_systems_stomp,Chatterjee:projection_based_look_ahead,Sundman:frogs,Needell:signal_recovery_from_incomplete_and_inaccurate_measurements_via_romp}. On the other hand, parallel types include \cosamp \cite{Needell:cosamp}, subspace pursuit (\sp)~\cite{Dai:subspace_pursuit} and their algorithmic variations \cite{Sundman:look_ahead_parallel_pursuit}. 
For any \cs reconstruction algorithm (convex optimization and \gp), providing theoretical reconstruction guarantees with relevant system requirements is a desired feature.
\cs for a single-sensor scenario including dynamic \cs \cite{Vaswani_KF_CS_2008,Carmi_KF_CS_2010,Zachariah_Chatterjee_Jansson_TSP_2012} has been substantially investigated in literature.

{{In this paper, we consider a distributed (or de-centralized) \cs (\dcs) problem where many sensors measure correlated sparse signals and the sensors are connected via a network. 
The task in a \dcs problem is reconstruction of correlated signals from the measurements collected in all sensor nodes.
\dcs has a wide range of application areas, 
for example, distributed sensor perception~\cite{Yang:distributed_perception} and distributed spectrum estimation~\cite{Feng:distributed_compressive_spectrum_sensing, Bazeraque:distributed_spectrum_sensing,Ling:decenteralized_support_detection,Sundman:psd}. Algorithms for the \dcs problem can be developed either in a central manner (by a fusion center) or distributed manner.
There are many centralized solutions, for example~\cite{Tropp:simultaneous_sparse_approx_part1,Rakotomamonjy:surveying,Leviatan:simultaneous,Cotter:sparse_solutions,Chen:theoretical_results_on_sparse,Sundman:greedy_pursuit_for_jointly} where measurements from all sensor nodes are collected in a central node and then the correlated sparse signals are reconstructed in the central node.
On the other hand, there are distributed algorithms where relevant information about correlations is exchanged over network and each sensor node reconstructs its own measured signal. Exchange of correlation information helps to improve quality of reconstruction in each sensor node.
A distributed algorithm is of high interest and we refer to such algorithms as \dcs algorithms. \dcs algorithms can be developed based on two principles: convex and \gp.
There is a tangible effort in the literature~\cite{Mota:distributed_basis_pursuit,Bazeraque:distributed_spectrum_sensing,Feng:distributed_compressive_spectrum_sensing,Ling:decenteralized_support_detection} to develop convex optimization based \dcs algorithms with theoretically proven convergence. The analytical tractability is due to use of convexity via implementation of distributed convex algorithms. 
For example, the work of~\cite{Mota:distributed_basis_pursuit} considered that a large measurement matrix in \cs is divided into many senors measuring one single source and the proposed algorithm is a distributed implementation of alternating-direction-method-of-multipliers (ADMM) \cite{Boyd_ADMM_2011}.
On the other hand, we note a limited endeavor to develop \dcs algorithms based on \gp principles. In this regard, our earlier attempts are in \cite{Sundman:diprsp,Sundman:a_greedy_pursuit_algorithm,Sundman:distributed_gp_algorithms} and some attempts by others are in \cite{Wimalajeewa:cooperative,Zhang:side_information_based_omp}. 
Most of these earlier attempts (including all our earlier attempts) were made for designing algorithms that can provide
a reasonable practical performance, but lacks theoretical guarantees on reconstruction performance. The relevant questions are: (a) what are system and signal properties so that a distributed algorithm converges, and (b) what is the quality of reconstruction performance at convergence? In fact, to the best of authors' knowledge, no significant theoretical results are available in current literature for \dcs algorithms based on \gp principles.
Naturally the limited endeavor may be attributed to the lack of analytical tractability.}}

{{We develop a new \dcs algorithm based on \gp principles that provide a good practical performance and have theoretical reconstruction guarantees.
Noting the important role of support-set estimation in \gp, sensor nodes exchange support set information over a network. 
For a signal model, we use the recent mixed support-set signal model of ~\cite{Sundman:a_greedy_pursuit_algorithm,Sundman:distributed_gp_algorithms} that considers correlation over support-sets of all underlying sparse signals in a \dcs problem. The correlation is incorporated via existence of a partial common (or joint) support-set; the common support-set is a subset of all individual supports of all sparse signals. Using the mixed support set model and appropriate assumptions about system setup, our contributions in this paper are:
\begin{itemize}
 \item Development of a distributed \gp algorithm.
 \item Analytical study of performance in the sense of provable reconstruction guarantees and convergence.
\end{itemize}
The new \dcs algorithm is referred to as \emph{distributed parallel pursuit} (\dipp) and it comprises of two main parts: data fusion and local \cs reconstruction. The task of the fusion is to provide an estimation of the correlation (i.e., estimation of the common support-set) which in turn helps to improve quality of the local \cs reconstruction.
For fusion, we use a democratic voting strategy. Typically a decision in voting strategy is made by majority counting (which may be considered a hard decision based approach), but the use of voting is motivated by simplicity and good performance to estimate the common support-set\footnote{We mention that a soft decision based approach does not suit well to estimate common support-set in our distributed setup where information about estimated support-sets for sensor nodes is exchanged over network.}.
The local \cs reconstruction algorithm use the output from the fusion as side information to improve reconstruction performance.
Based on the parallel pursuit algorithm \sp of~\cite{Dai:subspace_pursuit}, we design a new algorithm that can use the side information. The new algorithm is called \sipp (parallel pursuit with side information) and it is used as the local \cs re-constructor. While we develop \sipp by extending \sp, we could have used other parallel pursuit algorithms such as \cosamp \cite{Needell:cosamp} instead of \sp. The choice of parallel pursuit is due to algorithmic ease of incorporating side information and analytical tractability\footnote{In our distributed \cs setup we have found that the use of serial pursuit algorithms (such as \omp) comes with significant hurdles in analytical tractability and hence we do not explore serial pursuit algorithms in this paper.}.}}

{{\dipp works iteratively, where it improves the estimation of correlation by exchanging relevant information over the network.
Analysis of \dipp is non-trivial. In the literature, for analysis of \cs algorithms (or in general sparse representation problems), worst case analysis tools such as mutual-coherence and restricted-isometry-property (RIP) have been used \cite{Elad_book_2010}. The RIP was introduced in \cite{Candes:decoding_linear_programming} for analysis of convex optimization based \cs algorithms and later used significantly for analyzing \gp algorithms, such as \sp \cite{Dai:subspace_pursuit}. Instead of worst case approaches, average case analysis approaches such as replica method tools from statistical physics field \cite{Nishimori-2001} have been used for convex optimization based \cs algorithms \cite{Rangan_Replica_CS_TIT_2012,Kabashima_Vehkapera_Chatterjee_2012,Vehkapera_Kabashima_Chatterjee_TIT_2014}, but there is no precedence to use them for analyzing \gp algorithms due to analytical intractability.
Further, in worst case analysis tools, RIP is found to offer more analytical tractability than mutual-coherence and is recently in more use, such as analysis of \omp \cite{Davenport_2010_Orthogonal_Matching_pursuit}, model based \cs \cite{Baraniuk_Model_Based_CS_TIT_2010} and fusion framework where several \cs algorithms are used jointly \cite{Ambat:facs,Ambat_Chatterjee_Hari_2013_CoMACS}.  
Therefore we decide to use RIP to analyze \dipp. We show -- under certain theoretical requirements on the quality of information processing over network and RIP parameters of sensor nodes -- that the \dipp algorithm converges to a performance level (convergence in theory). At convergence, the performance level is a scaled additive measurement noise power where the scaling coefficient is a function of RIP parameters and information processing quality parameters. That means, under those theoretical requirements, the algorithm provides exact reconstruction if the measurement noise is absent. In practice, the algorithm iterates until the quality of correlation estimation saturates (convergence in practice) -- this happens when further information exchange does not help to improve reconstruction performance. Using simulations, we show how practical reconstruction performance of \dipp behaves with respect to change in number of measurements, signal-to-measurement-noise ratios and  network-connectivity conditions.}}

The remaining parts of the paper are organized as follows. In \sectionname~\ref{sec:setup}, we formally define the \dcs problem, the signal model and network models. Section~\ref{sec:algorithm} deals with developing \dipp. In \sectionname~\ref{sec:performance}, we derive performance bounds and reconstruction guarantees for \dipp. Lastly, in \sectionname~\ref{sec:numerical}, we perform practical evaluation of the \dipp algorithm by simulations.

\subsection{Notations and Preliminaries}
For enumerating sensor-nodes in the \dcs setup, we will reserve sub-indices `$\p$', `$\q$' and `$\r$'. However, to keep the paper clean from notational clutter, we will only use these sub-indices when it is necessary for the discussion. We reserve sub-indices `$l$' and `$k$' for denoting iteration counter in inner- and outer loops, respectively. Typically, `$l$' associates with the iteration counter of \sipp and `$k$' associates with \dipp.

Calligraphic letters are used for sets; in particular $\T$, $\J$ and $\I$ denote support-sets while $\mathcal{L}$ is a set of sensor nodes. We denote the full support set, $\Omega \triangleq \{1, 2, \dots, N \}$. Using $\Omega$, we define the complement $\T^{\complement} \triangleq \Omega \setminus \T$. If an algorithm at node $\p$ estimates the support-set, this estimate is denoted by $\Thi$. $\mbx_{\T}$ may refer to two things; either $\mbx_{\T}$ may be the non-zero sub-vector of $\mbx$ (i.e., $\mbx_{\T} = \{ x_i : i \in \T\}$), or $\mbx_{\T}$ may be a zero-padded signal, where $\mbx \in \mathbb{R}^N$, $\mbx_{\T} \neq 0$ and $\mbx_{\T^{\complement}} = 0$. Which one of these referred to will be clear from the context. When nothing else is stated the norm used is by default the induced $\ell_2$-norm (i.e., spectral norm) $\| \cdot \| \triangleq \| \cdot \|_2$. We define the pseudo-inverse for a matrix $\mbA$ as $\mbA^{\dagger} \triangleq (\mbA^*\mbA)^{-1}\mbA^*$ (where full column rank is assumed)
.

We now introduce some existing definitions and results for the standard \cs setup, that we will later use for the \dcs setup. In standard single-sensor \cs:
\begin{align}
\mathbf{y} = \mathbf{A} \mathbf{x} + \mathbf{e}, \label{eqn:single_sensor_model}
\end{align}
where $\mathbf{x} \in \mathbb{R}^{N}$ is a sparse signal, $\mathbf{y} \in \mathbb{R}^{M}$ is a measurement vector, $\mathbf{A} \in \mathbb{R}^{M\times N}$ is a measurement matrix and $\mathbf{e} \in \mathbb{R}^{M}$ is a measurement noise, and $M<N$.
\begin{definition}[RIP: Restricted Isometry Property \cite{Candes:decoding_linear_programming}]\label{def:rip}
A matrix $\mbA$ satisfies the RIP with Restricted Isometry Constant (RIC) $\delta_T$ if
\begin{align}
  (1-\delta_T)\|\mbx\|^2 \leq \| \mbA \mbx \|^2 \leq (1+\delta_T) \|\mbx\|^2
\end{align}
holds for all $T$-sparse vectors $\mbx$ where $0 \leq \delta_T < 1$.
\end{definition}

\begin{proposition}[Proposition 3.1 in \cite{Needell:cosamp}]\label{prop:cosamp}
  Suppose $\mbA$ has RIC $\delta_T$. Let $\T$ be a set of $T$ indices or fewer. Then
  \begin{subequations}
    \begin{align}
      \left\| \mbA_{\T}^* \mathbf{y} \right\| & \leq \sqrt{1+\delta_T}\left\| \mathbf{y} \right\|, \label{prop3.1:1} \\
      \left\| \mbA_{\T}^{\dagger} \mathbf{y} \right\| & \leq \frac{1}{\sqrt{1-\delta_T}} \left\| \mathbf{y} \right\|, \label{prop3.1:3}  \\
      (1-\delta_T) \left\| \mbx \right\| & \leq \left\| \mbA_{\T}^* \mbA_{\T} \mbx \right\| \leq (1+\delta_T) \left\| \mbx \right\|, \label{prop3.1:4} \\
      \frac{1}{1+\delta_T}\left\| \mbx \right\|\leq & \left\| (\mbA_{\T}^* \mbA_{\T})^{-1} \mbx \right\| \leq \frac{1}{1-\delta_T}\left\| \mbx \right\|. \label{prop3.1:5}
    \end{align}
  \end{subequations}
\end{proposition}
{{Here we mention that $\mbA_{\T}^*$ and $\mbA_{\T}^{\dagger}$ should be interpreted as $(\mbA_{\T})^*$ and $(\mbA_{\T})^{\dagger}$.}}
\begin{proposition}[Approximate Orthogonality] \textit{Proposition 3.2 in \cite{Needell:cosamp}} \label{prop:approx_orthogonality}.
Suppose $\mbA$ has RIC $\delta_T$. Let $\S$ and $\T$ be disjoint sets of indices whose combined cardinality does not exceed $S+T$. Then
\begin{align}
  \| \mbA_{\S}^* \mbA_{\T} \| \leq \delta_{S+T}.
\end{align}
\end{proposition}
\begin{corollary}[Corollary 3.3 in \cite{Needell:cosamp}] \label{cor:approx_orthogonality}
Suppose $\mbA$   has RIC $\delta_T$. Let $\T$ be a set of indices, and let $\mbx$ be a vector. Provided that $T \geq |\T \cup \supp(\mbx)|$,
\begin{align}
  \| \mbA_{\T}^* \mbA \mbx_{\T^{\complement}} \| = \| \mbA_{\T}^* \mbA_{{\T}^{\complement}} \mbx_{{\T}^{\complement}} \| \leq \delta_T\| \mbx_{{\T}^{\complement}} \|.
\end{align}
\end{corollary}
\begin{lemma}\label{lemma:bound_complement} For the setup \eqref{eqn:single_sensor_model}, if $\hat{\T}$ is the estimate of the support-set of a signal $\mbx$ and $\hat{\mbx}$ is constructed by $\hat{\mathbf{x}}_{\hat{\T}} \leftarrow \mathbf{A}^{\dagger}_{\hat{\T}} \mathbf{y}$, $\hat{\mathbf{x}}_{\hat{\T}^{\complement}} \leftarrow \mathbf{0}$, then the following relation holds:
\begin{align}
 \| \mbx - \hat{\mbx} \| \leq \frac{1}{1-\delta_{3T}} \| \mbx_{\hat{\T}^c} \| + \frac{1}{\sqrt{1-\delta_{3T}}} \|  \mathbf{e} \|. \label{lemma:compl1}
\end{align}
We also have that
  \begin{align}
    \| \mbx_{\hat{\T}^c} \| \leq \| \mbx - \hat{\mbx}\|. \label{lemma:compl2}
  \end{align}
\begin{IEEEproof} 
See \appendixname~\ref{app:lemma:bound_complement}. 
\end{IEEEproof}
\end{lemma}

\subsubsection{Some Algorithmic Notations}
For clarity in the algorithmic notations later, we define three algorithmic functions as follows:
\begin{align} 
{\supp}&(\mathbf{x}, k) \triangleq \{ \textit{the set of indices corresponding to} \nonumber \\
& \hspace{15mm}\textit{the $k$ largest amplitude components of } \mathbf{x} \}, \nonumber
\end{align}
and
\begin{align} 
\addi(\mathbf{s}, \T) & \triangleq \{ \forall j \in \T, \textit{ perform } s_{j} = s_{j}+1 \}, \nonumber 
\end{align}
where $\mathbf{s}=[s_{1} \,\, s_{2} \,\, \ldots s_{N} ]$ and $s_{j} \geq 0$. Lastly we define
\begin{align}
  \resid(\mby, \mbA) \triangleq \mby - \mbA\mbA^{\dagger}\mby,
\end{align}
for full column-rank matrices $\mathbf{A}$.

\section{Distributed Compressed Sensing Setup} \label{sec:setup}
The \dcs problem consists of several sensor nodes connected through a network, where the underlying data collected at the nodes are correlated. In this section we first describe the \dcs problem, then the correlation model and lastly we introduce the network model.

\subsection{Distributed Compressed Sensing}
In \dcs, the $\p$'th sensor measures a signal $\mathbf{x}_{\p} \in \mathbb{R}^{N}$ according to the following relation
\begin{align}
\mathbf{y}_{\p} = \mathbf{A}_{\p} \mathbf{x}_{\p} + \mathbf{e}_{\p}, ~~~~~~ \forall \p \in \mathcal{L}, \label{eqn:model}
\end{align}
where $\mathbf{y}_{\p} \in \mathbb{R}^{M}$ is the measurement vector, $\mathbf{A}_{\p} \in \mathbb{R}^{M\times N}$ is the measurement matrix, $\mathbf{e}_{\p} \in \mathbb{R}^{M}$ is the measurement noise and $\mathcal{L}$ is a global set containing all nodes in the network. Throughout this paper we use measurement matrices $\mathbf{A}_{\p}$ that have unit $\ell_2$-norm columns. This setup describes an under-determined system, where $M < N$. $\mathbf{A}_{\p}$ and $\mathbf{e}_{\p}$ are independent both locally and across the network. The signal vector $\mathbf{x}_{\p} = [x_{\p}(1) \,\, x_{\p}(2) \, \ldots \, x_{\p}(N)]$ is $T$-sparse, meaning it has $T$ elements that are non-zero. The element-indices corresponding to non-zero values are collected in the support-set $\Ti$, that means $\Ti = \{ i : x_{\p}(i) \neq 0 \}$ and $| \Ti | = T$. 
Next we discuss a relevant signal model that introduces correlation between $\{ \mathbf{x}_{\p} \}$.

\subsection{Correlation in Signals: Mixed Support-set Model} \label{sec:mixed_support_signal_model}
We introduce a mixed support-set signal model that brings correlation in signals through their support-sets. This model was previously presented in \cite{Sundman:greedy_pursuit_for_jointly} and \cite{Sundman:a_greedy_pursuit_algorithm}. For the sparse signal $\mathbf{x}_{\p}$, the support-set $\T_{\p}$ follows the construction
\begin{align}
\T_{\p} = \Ii \cup \J_{\p} = \Ii \cup \J, ~~~~~~ \forall \p \in \mathcal{L}. \label{eqn:support_set}
\end{align}
Here, the partial support-set $\J_{\p} =\J$ is joint (i.e., common) to the support-sets of all sparse signals, leading to correlation among signals $\{ \mathbf{x}_{\p} \}$. The other partial support-set $\Ii$ is individual and does not correspond to any correlation. 
\begin{assumption} Denoting $|\Ii|=I$ and $|\J|=J$, the following assumptions are used throughout the paper:
\begin{enumerate}
 \item Elements of support-sets are uniformly distributed.
 \item $\Ii \cap  \J = \emptyset, ~~~ \forall \p\in \mathcal{L}$.
 \item Hence, $T = I+J$.\wqed
\end{enumerate}
\end{assumption}
Additionally, we mention that no correlation between non-zero signal values of $\mathbf{x}_{\p}$ is assumed. The \dipp estimates $\J$ by cooperation (fusion) through relevant information exchange over a network and gradually improve \cs reconstruction performance at each sensor.  
  
We provide few examples of potential real life applications for the mixed support-set model: (1) spectrum estimation - where each node experiences large overlapping supports in the spectrum~\cite{Sundman:psd}, (2) multiple sensor image capturing - where each node observes the same object from slightly different angles~\cite{Kirmani:codac}, and (3) multiple sensor sound capturing~\cite{Wu:spherical_microphone}. Also for all the above scenarios and including the one-sensor scenario, if a slowly varying signal is tracked over time (dynamic \cs \cite{Vaswani_KF_CS_2008,Carmi_KF_CS_2010,Zachariah_Chatterjee_Jansson_TSP_2012}), the proposed mixed support-set model may also apply.

\subsection{Network Topology} \label{sec:network_model}

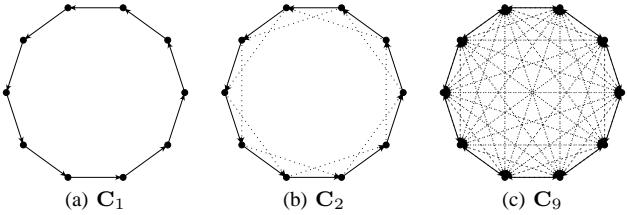
\begin{figure}[t]
  \linespread{0.5}
  \centering
  \subfloat[$\C_1$]{
    \resizebox{0.3\columnwidth}{!}{
      \begin{tikzpicture}[auto]
        \coordinate (a) at (0:15mm);
        \coordinate (b) at (1*36:15mm);
        \coordinate (c) at (2*36:15mm);
        \coordinate (d) at (3*36:15mm);
        \coordinate (e) at (4*36:15mm);
        \coordinate (f) at (5*36:15mm);
        \coordinate (g) at (6*36:15mm);
        \coordinate (h) at (7*36:15mm);
        \coordinate (i) at (8*36:15mm);
        \coordinate (j) at (9*36:15mm);
        \draw [fill] (a) circle (0.5mm);
        \draw [fill] (b) circle (0.5mm);
        \draw [fill] (c) circle (0.5mm);
        \draw [fill] (d) circle (0.5mm);
        \draw [fill] (e) circle (0.5mm);
        \draw [fill] (f) circle (0.5mm);
        \draw [fill] (g) circle (0.5mm);
        \draw [fill] (h) circle (0.5mm);
        \draw [fill] (i) circle (0.5mm);
        \draw [fill] (j) circle (0.5mm);
        \draw [line] (a) -> (b);
        \draw [line] (b) -> (c);
        \draw [line] (c) -> (d);
        \draw [line] (d) -> (e);
        \draw [line] (e) -> (f);
        \draw [line] (f) -> (g);
        \draw [line] (g) -> (h);
        \draw [line] (h) -> (i);
        \draw [line] (i) -> (j);
        \draw [line] (j) -> (a);
      \end{tikzpicture}
      \label{fig:network_1}
    }
  }
  \subfloat[$\C_2$]{
    \resizebox{0.3\columnwidth}{!}{
      \begin{tikzpicture}[auto]
        \draw [fill] (a) circle (0.5mm);
        \draw [fill] (b) circle (0.5mm);
        \draw [fill] (c) circle (0.5mm);
        \draw [fill] (d) circle (0.5mm);
        \draw [fill] (e) circle (0.5mm);
        \draw [fill] (f) circle (0.5mm);
        \draw [fill] (g) circle (0.5mm);
        \draw [fill] (h) circle (0.5mm);
        \draw [fill] (i) circle (0.5mm);
        \draw [fill] (j) circle (0.5mm);
       \draw [line] (a) -> (b);
        \draw [line] (b) -> (c);
        \draw [line] (c) -> (d);
        \draw [line] (d) -> (e);
        \draw [line] (e) -> (f);
        \draw [line] (f) -> (g);
        \draw [line] (g) -> (h);
        \draw [line] (h) -> (i);
        \draw [line] (i) -> (j);
        \draw [line] (j) -> (a);
        \draw [line,dotted] (a) -> (c);
        \draw [line,dotted] (b) -> (d);
        \draw [line,dotted] (c) -> (e);
        \draw [line,dotted] (d) -> (f);
        \draw [line,dotted] (e) -> (g);
        \draw [line,dotted] (f) -> (h);
        \draw [line,dotted] (g) -> (i);
        \draw [line,dotted] (h) -> (j);
        \draw [line,dotted] (i) -> (a);
        \draw [line,dotted] (j) -> (b);
      \end{tikzpicture}
      \label{fig:network_2}
    }
  }
  \subfloat[$\C_9$]{
    \resizebox{0.3\columnwidth}{!}{
      \begin{tikzpicture}[auto]
        \draw [fill] (a) circle (0.5mm);
        \draw [fill] (b) circle (0.5mm);
        \draw [fill] (c) circle (0.5mm);
        \draw [fill] (d) circle (0.5mm);
        \draw [fill] (e) circle (0.5mm);
        \draw [fill] (f) circle (0.5mm);
        \draw [fill] (g) circle (0.5mm);
        \draw [fill] (h) circle (0.5mm);
        \draw [fill] (i) circle (0.5mm);
        \draw [fill] (j) circle (0.5mm);
       \draw [line] (a) -> (b);
        \draw [line] (b) -> (c);
        \draw [line] (c) -> (d);
        \draw [line] (d) -> (e);
        \draw [line] (e) -> (f);
        \draw [line] (f) -> (g);
        \draw [line] (g) -> (h);
        \draw [line] (h) -> (i);
        \draw [line] (i) -> (j);
        \draw [line] (j) -> (a);
        \draw [line,dotted] (a) -> (c);
        \draw [line,dotted] (b) -> (d);
        \draw [line,dotted] (c) -> (e);
        \draw [line,dotted] (d) -> (f);
        \draw [line,dotted] (e) -> (g);
        \draw [line,dotted] (f) -> (h);
        \draw [line,dotted] (g) -> (i);
        \draw [line,dotted] (h) -> (j);
        \draw [line,dotted] (i) -> (a);
        \draw [line,dotted] (j) -> (b);

        \draw [line,dotted] (a) -> (d);
        \draw [line,dotted] (b) -> (e);
        \draw [line,dotted] (c) -> (f);
        \draw [line,dotted] (d) -> (g);
        \draw [line,dotted] (e) -> (h);
        \draw [line,dotted] (f) -> (i);
        \draw [line,dotted] (g) -> (j);
        \draw [line,dotted] (h) -> (a);
        \draw [line,dotted] (i) -> (b);
        \draw [line,dotted] (j) -> (c);

        \draw [line,dotted] (a) -> (e);
        \draw [line,dotted] (b) -> (f);
        \draw [line,dotted] (c) -> (g);
        \draw [line,dotted] (d) -> (h);
        \draw [line,dotted] (e) -> (i);
        \draw [line,dotted] (f) -> (j);
        \draw [line,dotted] (g) -> (a);
        \draw [line,dotted] (h) -> (b);
        \draw [line,dotted] (i) -> (c);
        \draw [line,dotted] (j) -> (d);

        \draw [line,dotted] (a) -> (f);
        \draw [line,dotted] (b) -> (g);
        \draw [line,dotted] (c) -> (h);
        \draw [line,dotted] (d) -> (i);
        \draw [line,dotted] (e) -> (j);
        \draw [line,dotted] (f) -> (a);
        \draw [line,dotted] (g) -> (b);
        \draw [line,dotted] (h) -> (c);
        \draw [line,dotted] (i) -> (d);
        \draw [line,dotted] (j) -> (e);

        \draw [line,dotted] (a) -> (g);
        \draw [line,dotted] (b) -> (h);
        \draw [line,dotted] (c) -> (i);
        \draw [line,dotted] (d) -> (j);
        \draw [line,dotted] (e) -> (a);
        \draw [line,dotted] (f) -> (b);
        \draw [line,dotted] (g) -> (c);
        \draw [line,dotted] (h) -> (d);
        \draw [line,dotted] (i) -> (e);
        \draw [line,dotted] (j) -> (f);

        \draw [line,dotted] (a) -> (h);
        \draw [line,dotted] (b) -> (i);
        \draw [line,dotted] (c) -> (j);
        \draw [line,dotted] (d) -> (a);
        \draw [line,dotted] (e) -> (b);
        \draw [line,dotted] (f) -> (c);
        \draw [line,dotted] (g) -> (d);
        \draw [line,dotted] (h) -> (e);
        \draw [line,dotted] (i) -> (f);
        \draw [line,dotted] (j) -> (g);

        \draw [line,dotted] (a) -> (i);
        \draw [line,dotted] (b) -> (j);
        \draw [line,dotted] (c) -> (a);
        \draw [line,dotted] (d) -> (b);
        \draw [line,dotted] (e) -> (c);
        \draw [line,dotted] (f) -> (d);
        \draw [line,dotted] (g) -> (e);
        \draw [line,dotted] (h) -> (f);
        \draw [line,dotted] (i) -> (g);
        \draw [line,dotted] (j) -> (h);

        \draw [line,dotted] (a) -> (j);
        \draw [line,dotted] (b) -> (a);
        \draw [line,dotted] (c) -> (b);
        \draw [line,dotted] (d) -> (c);
        \draw [line,dotted] (e) -> (d);
        \draw [line,dotted] (f) -> (e);
        \draw [line,dotted] (g) -> (f);
        \draw [line,dotted] (h) -> (g);
        \draw [line,dotted] (i) -> (h);
        \draw [line,dotted] (j) -> (i);
      \end{tikzpicture}
      \label{fig:network_3}
    }
  }
  \caption{Three different network topologies.}
  \label{fig:network}
\end{figure}

In \dcs, a sensor node is not aware of the full network topology. Instead, any node knows two sets of local neighbors; the incoming neighbor connections $\mathcal{L}_{\p}^{\text{in}}$ and outgoing neighbor connections $\mathcal{L}_{\p}^{\text{out}}$. Here \emph{incoming} and \emph{outgoing} connections corresponds to communication links where a node can receive or send information, respectively. 
{{In the paper, RIP based theoretical analysis of \dipp algorithm does not require a specific network topology (such as a bipartite graph used in \cite{Mota:distributed_basis_pursuit}) except the requirement that the given network is connected and static. 
To observe practical performance of \dipp via simulations we have considered two types of network: (a) structured, and (b) random. Let us describe the first type (structured network).
}}
Using the two sets $\mathcal{L}_{\p}^{\text{in}}$ and $\mathcal{L}_{\p}^{\text{out}}$, consider a number of nodes topologically arranged in a circle. By letting each node forwardly connect to one other node (i.e., node $\p$ get $\mathcal{L}_{\p}^{\text{in}} = \{ \p-1 \}$ and $\mathcal{L}_{\p}^{\text{out}} = \{ \p+1 \}$), a circular topology can be created; we refer to such network as a degree-one network topology. Using ten nodes, we denote the degree-one network by a connection-matrix $\C_1$, depicted in \figurename~\ref{fig:network_1}. A degree-two $\C_2$ network is shown in \figurename~\ref{fig:network_2} and 
a degree-nine $\C_9$ network is shown in \figurename~\ref{fig:network_3}. 
We use this structured network topology so that improvement in CS reconstruction performance of \dipp vis-a-vis increase in network connection can be studied in a controlled manner. 
{{
Next, for the second type (random network), we considered Watts-Strogatz network model~\cite{watts1998collective} that is claimed to have many practically relevant applications.  The Watts-Strogatz network model typically considers a large number of nodes, and has two parameters $q$ and $p$.
Using these parameters, first, every node gets connected to $q$ neighbors in a structural manner via bi-directional communication links. 
Then, every connection is rewired with probability $p$ to another node chosen uniformly at random.
}}


\section{Distributed Parallel Pursuit: Algorithm} \label{sec:algorithm}

\begin{figure}[t]
  \tikzstyle{method}=[draw,rectangle,inner sep=2pt,minimum height=10mm,thin]
  \tikzstyle{label}=[yshift={(-1mm)}]
  \pgfdeclarelayer{background}
  \pgfdeclarelayer{foreground}
  \pgfsetlayers{background,main,foreground} 
  \centering
  \begin{tikzpicture}[auto]
    \footnotesize\sffamily
   \begin{pgfonlayer}{foreground}
    \node[method,minimum width=10mm,minimum width=20mm] (ipp) at (0,0) {\normalsize\sipp};
    \node[method,dashed,minimum width=10mm,rounded corners] (nout) at (43mm,0) {$\mathcal{L}^{\text{out}}_{\p}$};
    \node[method,dashed,minimum width=10mm,rounded corners] (nin) at (43mm,-20mm) {$\mathcal{L}^{\text{in}}_{\p}$};
    \node[method,minimum width=10mm,fill=white] (con) at (12mm,-20mm) {\footnotesize\consensus};
    \node[method,minimum width=10mm,fill=white] (exp) at (-12mm,-20mm) {\footnotesize\expansion};
    \end{pgfonlayer}


    \draw[dashed,thin,rounded corners]
    ([yshift=4mm,xshift=-4mm] exp.north west) --
    node[pos=0.15, below] {Fusion}
    ([yshift=4mm,xshift=10mm] con.north east) --
    ([yshift=-2mm,xshift=10mm] con.south east) --
    ([yshift=-2mm,xshift=-4mm] exp.south west) -- cycle;

    \draw[dashed,thin,rounded corners] 
    ([yshift=4mm,xshift=-2mm] nout.north west) --
    node[midway, below] {Network}
    ([yshift=4mm,xshift=2mm] nout.north east) --
    ([yshift=-4mm,xshift=2mm] nin.south east) --
    ([yshift=-4mm,xshift=-2mm] nin.south west) -- cycle;

    \gettikzxy{(ipp.north)}{\junk}{\ippy}
    \gettikzxy{(exp.west)}{\expx}{\junk}
    \gettikzxy{(con.east)}{\conx}{\junk}
    \gettikzxy{(con.south)}{\junk}{\cony}
    \begin{pgfonlayer}{background}
      \draw[thin]
      ([yshift=4mm,xshift=-6mm] \expx, \ippy) --
      node[pos=0.07, below] {\normalsize\dipp}
      ([yshift=4mm,xshift=12mm] \conx, \ippy) --
      ([yshift=-4mm,xshift=12mm] \conx, \cony) --
      ([yshift=-4mm,xshift=-6mm] \expx, \cony) -- cycle;
    \end{pgfonlayer}

    \draw[dspconn,thin] (ipp.east) -- node (conn1) {} node[label,midway] {$\Th_{\p}$} (nout.west);
    \draw[dspconn,thin] ($(nin.south west)!0.25!(nin.north west)$) -- node[label,above,pos=0.65] {$\{\Th_{\q}\}$}
    ($(con.south east)!0.25!(con.north east)$);
    \draw[dspconn,thin] ($(con.south west)!0.25!(con.north west)$) -- node[label,above,midway] {$\hat{\J}_{\p}$}
    ($(exp.south east)!0.25!(exp.north east)$);

    \gettikzxy{($(con.south east)!0.75!(con.north east)$)}{\linex}{\liney}    
    \gettikzxy{(conn1)}{\connex}{\conney}    
    \draw[thin] (conn1) -- node[midway] (conn2) {} (\connex,\liney);
    \draw[dspconn,thin] (\connex,\liney) -- ($(con.south east)!0.75!(con.north east)$);

    \gettikzxy{($(exp.south east)!0.75!(exp.north east)$)}{\ex}{\ey}    
    \gettikzxy{(conn2.west)}{\cx}{\cy}    
    \draw[thin] ([yshift=-5mm] \cx,\cy) node[dspnodefull,thin] {} -- ([yshift=-5mm,xshift=4mm] \ex,\cy) -- ([xshift=4mm] \ex,\ey);
    \draw[dspconn,thin] ([xshift=4mm] \ex,\ey) -- (\ex,\ey);

    \gettikzxy{(exp.west)}{\ex}{\ey}    
    \gettikzxy{($(ipp.south west)!0.25!(ipp.north west)$)}{\ix}{\iy}
    \draw[thin] (\ex,\ey) -- ([xshift=-2mm] \ex,\ey) -- ([xshift=-2mm] \ex,\iy);
    \draw[dspconn,thin] ([xshift=-2mm] \ex,\iy) -- node[label,midway] {$\T_{\p,\text{si}}$} (\ix,\iy);

    \gettikzxy{($(ipp.south west)!0.75!(ipp.north west)$)}{\ix}{\iy}
    \draw[dspconn,thin] ([xshift=-30mm] \ix,\iy) -- node[label,pos=0.24] {$\mathbf{y}_{\p}, \mathbf{A}_{\p}, T$} (\ix,\iy);


  \end{tikzpicture}
  \caption{Distributed parallel pursuit (\dipp) and the network.}\label{fig:dipp}
\end{figure}
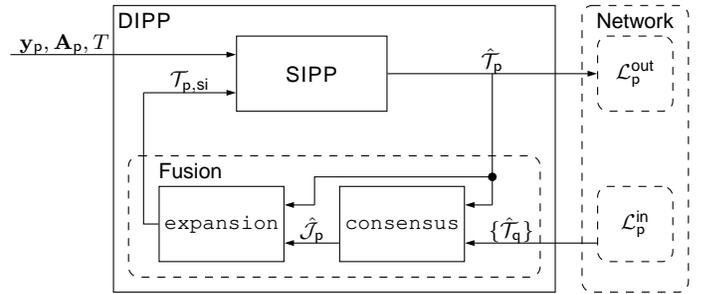

{{
Considering the importance of accurate support-set estimation in \gp algorithms, we endeavor to develop a distributed \gp by considering exchange (or communication) of support-set information over the network. By fusing the support-set information and providing the result as side information to \sipp we develop the distributed parallel pursuit (\dipp) algorithm.
A block diagram of \dipp is shown in \figurename~\ref{fig:dipp}. \dipp is executed in each node and it comprises of two main parts: (1) a \cs reconstruction algorithm - \sipp, and (2) fusion of estimated support-sets. The fusion comprises of two sub-algorithms: (a) a consensus strategy by voting, and (b) an expansion strategy. All parts in \dipp are developed to bring a suitable balance between practical engineering and analytical tractability. With our objective of theoretical convergence, we take a strategy in designing \sipp such that it has a proven convergence and also can use a side information. As \sipp is a part of \dipp, the convergence of \sipp helps to prove the convergence of \dipp. Further, the consensus and expansion sub-algorithms in the fusion block help to find relevant side information in an efficient manner that leads to a better practical performance and analytical tractability for \dipp. We describe these algorithmic parts one-by-one in the following 
subsections, and later analyze them in section~\ref{sec:performance}. 
}}

\subsection{SIPP: Parallel Pursuit with Side Information} \label{sec:alg:pp}

{{
The existing \sp algorithm~\cite{Dai:subspace_pursuit} is appropriately modified to develop \sipp, described in \algorithmname~\ref{alg:sipp}. Compared to \sp, the modifications are: the supply of $\T_{{\p},\text{si}}$ as side information satisfying $|\T_{{\p},\text{si}}| = T$, and the addition of steps \ref{alg:pp:t_acute} to \ref{alg:pp:x_hat} for using the side information to achieve a better local estimation of $\hat{\mathbf{x}}_{\p}$ and $\hat{\T}_{\p}$. Note the step \ref{alg:pp:u_check} where $\T_{\text{si}} = \T_{{\p},\text{si}}$ participates in \cs reconstruction of \sipp.
}}
As a \textit{stopping criterion}, we can use an upper limit of allowable iterations and/or violation of a non-decreasing residual norm condition, that is if $\| \mathbf{r}_l \| \leq \| \mathbf{r}_{l-1} \|$ violates. Finally, note that if $\T_{{\p},\text{si}} = \emptyset$ then $\tilde{\mathbf{x}}_l = \hat{\mathbf{x}}_l$, and the \sipp becomes identical to \sp~\cite{Dai:subspace_pursuit}. 
{{In step \ref{alg:pp:x_tilde} of algorithm \ref{alg:sipp}, for notational clarity, we use the notation $\tilde{\mathbf{x}}_{\tilde{\mathcal{U}}_l} \triangleq (\tilde{\mathbf{x}}_l)_{\tilde{\mathcal{U}}_l}$ to denote the coordinates of $\tilde{\mathbf{x}}_l$ which are indexed by the set $\tilde{\mathcal{U}}_l$. Similarly in step \ref{alg:pp:x_check} we use the notation $\check{\mathbf{x}}_{\check{\mathcal{U}}_l} \triangleq (\check{\mathbf{x}}_l)_{\check{\mathcal{U}}_l}$ to denote the coordinates of $\check{\mathbf{x}}_l$ which are indexed by the set $\check{\mathcal{U}}_l$; the same strategy for notation is also used in step \ref{alg:pp:x_hat} for $\hat{\mathbf{x}}_l$.
}}

\begin{algorithm}
\caption{\sipp (parallel pursuit with side information): \textit{Executed in the local node $\p$} } \label{alg:sipp}
\textit{Input:} $\mathbf{y}_{\p}$, $\mathbf{A}_{\p}$, $T$, $\T_{{\p},\text{si}}$ \\
\textit{Initialization:} 
\begin{algorithmic}[1]
\STATE $\mathbf{y} \leftarrow \mathbf{y}_{\p}$, $\mathbf{A} \leftarrow \mathbf{A}_{\p}$, $\T_{\text{si}} \leftarrow \T_{{\p},\text{si}}$
\STATE$l \leftarrow 0$, $\mathbf{r}_l \leftarrow \mathbf{y}$, $\hat{\T}_l \leftarrow \emptyset$, $\hat{\mathbf{x}}_l \leftarrow 0$ 
\end{algorithmic}
\textit{Intermediate variables:}
\begin{algorithmic}[1]
\STATE $\tilde{\mathbf{x}}_l \in \mathbb{R}^{N}$, $\check{\mathbf{x}}_l \in \mathbb{R}^{N}$, $\hat{\mathbf{x}}_l \in \mathbb{R}^{N}$
\end{algorithmic}
\textit{Iteration:}
\begin{algorithmic}[1]
\REPEAT
\STATE $l \leftarrow l + 1$ \hfill (Iteration counter)
\STATE $\grave{\T}_l \leftarrow \supp(\mathbf{A}^*\mathbf{r}_{l-1}, T)$ \label{alg:pp:grave}
\STATE $\tilde{\mathcal{U}}_l \leftarrow \grave{\T}_l \cup \hat{\T}_{l-1}$ \label{alg:pp:u_tilde}
\STATE $\tilde{\mathbf{x}}_l \leftarrow \mathbf{0}$; $\tilde{\mathbf{x}}_{\tilde{\mathcal{U}}_l} \leftarrow \mathbf{A}^{\dagger}_{\tilde{\mathcal{U}}_l} \mathbf{y}$ \hfill (Note: $\tilde{\mathbf{x}}_{\tilde{\mathcal{U}}_l^{\complement}} = \mathbf{0}$) \label{alg:pp:x_tilde}
\STATE $\acute{\T}_l \leftarrow \supp(\tilde{\mathbf{x}}_l, T)$ \label{alg:pp:t_acute}
\STATE $\check{\mathcal{U}}_l \leftarrow \acute{\T}_l \cup \T_{\text{si}}$ \label{alg:pp:u_check}
\STATE $\check{\mathbf{x}}_l \leftarrow \mathbf{0}$; $\check{\mathbf{x}}_{\check{\mathcal{U}}_l} \leftarrow \mathbf{A}^{\dagger}_{\check{\mathcal{U}}_l} \mathbf{y}$ \hfill (Note: $\check{\mathbf{x}}_{\check{\mathcal{U}_l}^{\complement}} = \mathbf{0}$) \label{alg:pp:x_check}
\STATE $\hat{\T}_l \leftarrow \supp(\check{\mathbf{x}}_l, T)$ \label{alg:pp:t}
\STATE $\hat{\mathbf{x}}_l \leftarrow \mathbf{0}$; $\hat{\mathbf{x}}_{\hat{\T}_l} \leftarrow 
\mathbf{A}^{\dagger}_{\hat{\T}_l} \mathbf{y}$ \hfill (Note $\hat{\mathbf{x}}_{\hat{\T}_l^{\complement}} = 
\mathbf{0}$) \label{alg:pp:x_hat}
\STATE $\mathbf{r}_l \leftarrow \mathbf{y} - \mathbf{A}\hat{\mathbf{x}}_l$ \label{alg:pp:r}
\UNTIL \textit{stopping criterion}
\end{algorithmic}
\textit{Output:} ~ $ \hat{\mathbf{x}}_{\p} \leftarrow \hat{\mbx}_{l}$, $\hat{\T}_{\p} \leftarrow \hat{\T}_{l}$, $\mathbf{r}_{\p} \leftarrow \mathbf{r}_{l}$
\end{algorithm}

In \dipp, after execution of \sipp, the support-set estimate $\hat{\T}_{\p}$ is broadcasted over the network and later fused to estimate common support-set ${\J}_{\p} = \J$ in each node.


\subsection{Fusion} \label{sec:alg:consensus}

Our fusion strategy is presented in algorithm~\ref{alg:fusion} that comprises of two sub-algorithms: \consensus and \expansion. The $\p$'th node has access to support-set estimates $\{ \hat{\T_{\q}} \}_{\q \in \mathcal{L}_{\p}^{\text{in}}}$ from neighbors, and the local estimate $\hat{\T}_{\p}$ and $\hat{\mbx}_{\p}$ (provided by the local \sipp algorithm). Based on this, the task of the \consensus algorithm is to estimate the common support-set as $\hat{\J}_{\p}$ such that $| \hat{\J}_{\p} | \leq T$ and the \expansion expands $\hat{\J}_{\p}$ to final output of fusion as the side-information $\T_{{\p},\text{si}}$ such that $|\T_{{\p},\text{si}}| = T$. 


{{
The \consensus strategy is to choose those indices for $\Jh_{\p}$ that are present in support-sets of at least two incoming neighboring nodes.  Studying \algorithmname~\ref{alg:consensus}, the inputs are: a set of estimated support-sets $\{ \hat{\T}_{\q} \}_{\q \in \mathcal{L}^{\text{in}}}$ from the neighbors, the own estimated support-set $\hat{\T}_{\p}$ and signal $\hat{\mbx}_{\p}$, and the sparsity level $T$. The estimate of $\hat{\J}_{\p}$ is formed (step~\ref{alg:cons:j}) such that no index in $\hat{\J}_{\p}$ has less than two votes (i.e., each index in $\hat{\J}_{\p}$ is present in at least two support-sets from $\{ \{ \hat{\T}_{\q} \}_{\q \in \mathcal{L}_{\p}^{\text{in}}}, \hat{\T}_{\p}\}$)\footnote{For node $\p$, this is equivalent to let 
algorithm choose $\Jh_{\p}$ as the union of all pair-wise intersections of support-sets (see the analysis section~\ref{sec:consensus} for details).
}. If the number of indices with at least two votes exceed the cardinality $T$, we pick the elements of $\hat{\J}_{\p}$ lexicographically. Our assumption is that an index present in two nodes' support-set estimates has a high probability of being in the common support $\J$. The assumption is based on a standard democratic voting principle where majority based decision is typically honoured.
A natural question is why we use a voting based consensus, but not a soft-decision based approach? 
The answer lies in a practical and inherent engineering aspect that an element of a support-set can be found in several estimated support-sets, and the decision to include such an element to be part of common support-set $\J$ can be efficiently done by voting based consensus (a counting process and seeking majority). On the other hand, a soft-decision based approach require explicit design with optimality conditions.
}}

{{
Next, in the \expansion sub-algorithm, the task is to expand $\hat{\J}_{\p}$ to $\T_{{\p},\text{si}}$ such that $|\T_{{\p},\text{si}}| = T$, which is later used as a side information. Note that $\hat{\I}_{\p} \subset \hat{\T}_{\p} =  \supp(\hat{\mbx}_{\p})$ and $\Thi \setminus \hat{\I}_{\p}$ does not contribute to form $\T_{{\p},\text{si}}$. Further note that $\hat{\J}_{\p} \subset \T_{{\p},\text{si}}$, that means we trust $\hat{\J}_{\p}$ in an absolute manner. It can be safely assumed that the estimation quality of $\hat{\J}_{\p}$ increases with the increase in quality of information exchange over network; for example, increase in network connectivity results in more incoming neighbor connections indexed by $\mathcal{L}_{\p}^{\text{in}}$, which in turn leads to a better voting in \consensus and a better output of fusion $\T_{{\p},\text{si}}$.
}}

{{
\begin{algorithm}[ht!]
\caption{{{Fusion: \textit{Executed in the local node $\p$}}}} \label{alg:fusion}
\textit{Fusion comprises of two sub-algorithms: \consensus and \expansion} \hrule  
\textit{Input:} $\{ \hat{\T}_{\q} \}_{\q \in \mathcal{L}_{\p}^{\text{in}}}$, $\hat{\T}_{\p}$, $T$, $\hat{\mbx}_{\p}$ \\
\textit{Initialization:} $\mathbf{z} \leftarrow \mathbf{0}_{N\times 1}$ \label{alg:cons:0} \\
\textit{\consensus sub-algorithm:} \label{alg:consensus}
\begin{algorithmic}[1]
\STATE $\mathbf{z} \leftarrow \addi(\mathbf{z}, \hat{\T}_{\p})$  \hspace{1cm} ($\p$-th node's estimate) \label{alg:cons:own}
\FOR { each $\q \in \mathcal{L}_{\p}^{\text{in}}$}
\STATE $\mathbf{z} \leftarrow \addi(\mathbf{z}, \hat{\T}_{\q})$ \hspace{1cm} (The neigbors' estimates) \label{alg:cons:others}
\ENDFOR
\STATE Choose $\hat{\J}_{\p}$ ~ s.t. ~ $(z(i) \geq 2) ~ \forall i \in \Jh_{\p}$ and $| \Jh_{\p} | \leq T$ \label{alg:cons:j}
\end{algorithmic}
\mbox{\textit{\consensus output:} $\Jh_{\p}$} \\
\textit{\expansion sub-algorithm:}
\begin{algorithmic}[1]
\STATE $(\hat{\mbx}_{\p})_{\hat{\J}_{\p}} \leftarrow \mathbf{0}$
\STATE $\hat{\I}_{\p} \leftarrow \supp(\hat{\mbx}_{\p}, T - |\hat{\J}_{\p}|)$
\end{algorithmic}
\textit{\expansion output:} $\T_{{\p},\text{si}} \leftarrow \hat{\I}_{\p} \cup \hat{\J}_{\p}$  \hfill ($|\T_{{\p},\text{si}}| = T$) \\
\textit{Final output of fusion:} $\T_{{\p},\text{si}}$, $\hat{\J}_{\p}$
\end{algorithm}
}}

\subsection{DIPP: Distributed Parallel Pursuit} \label{sec:alg_dpp}
Using \algorithmname~\ref{alg:sipp} and \ref{alg:fusion}, we now develop the distributed parallel pursuit (\dipp) presented in \algorithmname~\ref{alg:dipp}.
Input to \algorithmname~\ref{alg:dipp} for the $\p$'th node is the measurement signal $\mathbf{y}_{\p}$, the measurement matrix $\mbA_{\p}$, and sparsity $T$. Also \algorithmname~\ref{alg:dipp} knows $\mathcal{L}_{\p}^{\text{in}}$ and $\mathcal{L}_{\p}^{\text{out}}$. We assume that some underlying communication scheme provides for the transmit and receive functionality. In the initialization phase, an iteration parameter `$k$' is set to $0$ and the \sipp algorithm is executed with $\T_{\p,\text{si}} = \emptyset$.

\begin{algorithm}
\caption{Distributed parallel pursuit (\dipp) \newline \textit{Executed in the local node $\p$}} \label{alg:dipp}
\textit{Input:} $\mathbf{y}_{\p}$, $\mathbf{A}_{\p}$, $T$, $\mathcal{L}_{\p}^{\text{in}}$, $\mathcal{L}_{\p}^{\text{out}}$ \\
\textit{Initialization:} $k \leftarrow 0$, $(\hat{\mbx}_{\p,k}, \hat{\T}_{\p,k}, \mathbf{r}_{\p,k}) \leftarrow \sipp(\mathbf{y}_{\p}, \mbA_{\p}, T, \emptyset)$ \\
\textit{Iteration:}
\begin{algorithmic}[1]
\REPEAT
\STATE $k \leftarrow k + 1$  \hfill (Iteration counter)
\STATE \textbf{Transmit:} $\hat{\T}_{\p,k}$ to all nodes $\p \in \mathcal{L}_{\p}^{\text{out}}$ \label{alg:dpp:transmit}
\STATE \textbf{Receive:} $\hat{\T}_{\q,k}$ from all nodes $\q \in \mathcal{L}_{\p}^{\text{in}}$ \label{alg:dpp:receive}
\STATE $\hat{\J}_{\p,k} \leftarrow \consensus(\{ \hat{\T}_{\q,k} \}_{\q \in \mathcal{L}_{\p}^{\text{in}}}, \hat{\T}_{\p,k-1} ,T)$ \label{alg:dpp:consensus}
\STATE $\T_{\p,\text{si},k} \leftarrow \expansion(\hat{\J}_{\p,k}, \hat{\mbx}_{\p,k-1}, T)$ \label{alg:dpp:merger}
\STATE $(\hat{\mbx}_{\p,k}, \hat{\T}_{\p,k}, \mathbf{r}_{\p,k}) \leftarrow \sipp(\mathbf{y}_{\p}, \mbA_{\p}, T, \T_{\p,\text{si},k})$
\UNTIL \textit{stopping criterion}
\end{algorithmic}
\textit{Output:} ~ $\hat{\mbx}_{\dipp} \leftarrow \hat{\mbx}_{\p, k}$, ~ $\hat{\T}_{\dipp} \leftarrow \hat{\T}_{\p, k}$
\end{algorithm}

In the iterations, support-set estimates are exchanged over the network (steps \ref{alg:dpp:transmit} and \ref{alg:dpp:receive}). For the $k$'th iteration, the \consensus algorithm produces an estimate of the common support-set $\hat{\J}_{\p,k}$ (step \ref{alg:dpp:consensus}). Then, the \expansion (step \ref{alg:dpp:merger}) is used to extend $\hat{\J}_{\p,k}$, using $\hat{\T}_{\p,k-1}$, to produce $\T_{\p,\text{si},k}$. $\T_{\p,\text{si},k}$ is then used as an input to \sipp for an updated estimation of signal as $\hat{\mbx}_{\p,k}$ and support-set $\hat{\T}_{\p,k}$. As \textit{stopping criterion}, we can use an upper limit of iterations and/or violation of non-decreasing residual norm condition (i.e., violation of $\| \mathbf{r}_{\p,k} \| \leq \| \mathbf{r}_{\p,k-1} \|$).

\section{Distributed Parallel Pursuit: Analysis} \label{sec:performance}

In this section we will provide theoretical reconstruction guarantees with corresponding system requirements for the \dipp algorithm. We first analyze \sipp, \consensus and \expansion separately, and then provide the analysis for \dipp.


\subsection{Analysis of \sipp}
In this section, we will derive the reconstruction guarantee of \sipp. As \sipp is executed in each node $\p$, we drop the index $\p$ to avoid a notational clutter. The reconstruction guarantee is presented as a performance bound in \propositionname~\ref{thm:perf_bound_pp}. To derive the bound, we recursively apply a recurrence inequality which is shown in \propositionname~\ref{thrm:recurrence_ipp}. The recurrence inequality describes the change in reconstruction quality between iteration $l$ and $l-1$. We first introduce following lemmas.

\begin{lemma} \label{lemma:step1}
\begin{align}
\| \mbx_{\hat{\T}_l^{\complement}} \| \leq  \frac{1+\delta_{3T}}{1-\delta_{3T}} \| \mbx_{\check{\mathcal{U}}_l^{\complement}} \| + \frac{2}{\sqrt{1- \delta_{3T}}} \| \mathbf{e} \| \nonumber
\end{align}
\begin{IEEEproof}
  see \appendixname~\ref{app:lemma:step1}.
\end{IEEEproof}
\end{lemma}

\begin{lemma} \label{lemma:sp2}
  \begin{align}
    \| \mbx_{\acute{\T}_l^{\complement}} \| \leq \frac{1+\delta_{3T}}{1-\delta_{3T}} \| \mbx_{\tilde{\mathcal{U}}_l^{\complement}} \| + \frac{2}{\sqrt{1-\delta_{3T}}} \| \mathbf{e} \| \nonumber
\end{align}
\begin{IEEEproof}
By studying \algorithmname~\ref{alg:sipp}, it is clear that the functionality between steps \ref{alg:pp:u_tilde} and \ref{alg:pp:t_acute} are the same as between steps \ref{alg:pp:u_check} and \ref{alg:pp:t}. Thus, by replacing $\Th_l$ with $\acute{\T}_l$, $\check{\mathcal{U}_l}$ with $\tilde{\mathcal{U}}_l$ and $\check{\mbx}_l$ with $\tilde{\mbx}_l$ in the proof of \lemmaname~\ref{lemma:step1}, we arrive at this inequality.
\end{IEEEproof}
\end{lemma}

\begin{lemma} \label{lemma:sp1}
  \begin{align}
  \| \mbx_{\tilde{\mathcal{U}}_l^{\complement}} \| \leq \frac{2\delta_{3T}}{(1-\delta_{3T})^2} \| \mbx_{\Th_{l-1}^{\complement}} \| + \frac{2\sqrt{1+\delta_{3T}}}{1-\delta_{3T}} \| \mathbf{e} \| \nonumber
\end{align}
\begin{IEEEproof}
  see \appendixname~\ref{app:lemma:sp1}.
\end{IEEEproof}
\end{lemma}

Using these lemmas we are now ready to derive the recurrence inequality of \sipp.

\begin{proposition}[Recurrence inequality of \sipp] \label{thrm:recurrence_ipp}
\begin{align}
  \| \mbx_{\hat{\T}_l^{\complement}} \| & \leq a_{\sipp} \|\mbx_{\hat{\T}_{l-1}^{\complement}} \|  + b_{\sipp} \| \mbx_{\T_\text{si}^{\complement}}\| + c_{\sipp} \| \mathbf{e} \|, \nonumber
\end{align}
where
\begin{align}
  a_{\sipp} & \triangleq \frac{\delta_{3T} (1+\delta_{3T})^2}{(1-\delta_{3T})^4} {{>0}} \nonumber \\
  b_{\sipp} & \triangleq \frac{1 + \delta_{3T}}{2(1-\delta_{3T})} {{>0}} \nonumber  \\
  c_{\sipp} & \triangleq \frac{4(1+\delta_{3T}^2)}{(1-\delta_{3T})^3} {{>0}}. \nonumber 
\end{align}
\begin{IEEEproof}
  To prove the recurrence inequality of \sipp, we apply inequalities tracing backwards for sub-parts of \algorithmname~\ref{alg:sipp}. We will apply the inequalities in the following order:
  \begin{enumerate}
  \item Steps~\ref{alg:pp:t} to \ref{alg:pp:u_check} by using \lemmaname~\ref{lemma:step1}.
  \item Steps~\ref{alg:pp:u_check} to \ref{alg:pp:t_acute} by forming a new inequality.
  \item Steps~\ref{alg:pp:t_acute} to \ref{alg:pp:u_tilde} by using \lemmaname~\ref{lemma:sp2}.
  \item Steps~\ref{alg:pp:u_tilde} (of iteration $l$) to \ref{alg:pp:t} (of iteration $l-1$) by using \lemmaname~\ref{lemma:sp1}.
  \end{enumerate}
  
  By combining $\|
  \mbx_{\check{\mathcal{U}}_l^{\complement}} \| \leq \|  \mbx_{\acute{\T}_l^{\complement}} \|$ and $\|
  \mbx_{\check{\mathcal{U}}_l^{\complement}} \| \leq \|
  \mbx_{\T_{\text{si}}^{\complement}} \|$, we can write
  \begin{align} \label{eqn:initial_ineq} \|
    \mbx_{\check{\mathcal{U}}_l^{\complement}} \| \leq \frac{\|
      \mbx_{\acute{\T}_l^{\complement}} \| + \|
      \mbx_{\T_{\text{si}}^{\complement}} \|}{2}.
  \end{align}
  By using \eqref{eqn:initial_ineq} in lemma~\ref{lemma:step1}, we get
  \begin{align}
    & \| \mbx_{\hat{\T}_l^{\complement}} \| \leq  \frac{1+\delta_{3T}}{1-\delta_{3T}} \| \mbx_{\check{\mathcal{U}_l}^{\complement}} \| + \frac{2}{\sqrt{1- \delta_{3T}}} \| \mathbf{e} \| \nonumber \\
    & \phantom{=} \leq \frac{1+\delta_{3T}}{2(1-\delta_{3T})} \| \mbx_{\acute{\T_l}^{\complement}} \| + \frac{1+\delta_{3T}}{2(1-\delta_{3T})} \| \mbx_{\T_{\text{si}}^{\complement}} \| + \frac{2}{\sqrt{1- \delta_{3T}}} \| \mathbf{e} \|. \label{steps1-2}
  \end{align}
  We now apply \lemmaname~\ref{lemma:sp2} to \eqref{steps1-2}
  \begin{align}
    & \| \mbx_{\hat{\T}_l^{\complement}} \| \leq \frac{1+\delta_{3T}}{2(1-\delta_{3T})} \left( \frac{1+\delta_{3T}}{1-\delta_{3T}} \| \mbx_{\tilde{\mathcal{U}}_l^{\complement}} \| + \frac{2}{\sqrt{1-\delta_{3T}}} \| \mathbf{e} \| \right) \nonumber \\
    & \phantom{=} \phantom{=} + \frac{1+\delta_{3T}}{2(1-\delta_{3T})} \| \mbx_{\T_{\text{si}}^{\complement}} \| + \frac{2}{\sqrt{1- \delta_{3T}}} \| \mathbf{e} \| \nonumber \\
    & \phantom{=} \overset{(a)}{\leq} \frac{(1\!+\!\delta_{3T})^2}{2(1\!-\!\delta_{3T})^2}\! \| \mbx_{\tilde{\mathcal{U}}_l^{\complement}} \| + \frac{1\!+\!\delta_{3T}}{2(1\!-\!\delta_{3T})}\! \| \mbx_{\T_{\text{si}}^{\complement}} \| + \frac{3\!-\!\delta_{3T}}{(1\!-\!\delta_{3T})^2}\! \| \mbe \|. \label{step2-3}
  \end{align}
  Here, we have in \ensuremath{(a)} used that $1-\delta_{3T} \leq \sqrt{1-\delta_{3T}}$. To finalize the bound we apply \lemmaname~\ref{lemma:sp1} to \eqref{step2-3}
  \begin{align}
    & \| \mbx_{\hat{\T}_l^{\complement}} \| \leq \frac{(1+\delta_{3T})^2}{2(1-\delta_{3T})^2} \left(\!\! \frac{2\delta_{3T}}{(1-\delta_{3T})^2} \| \mbx_{\Th_{l-1}^{\complement}} \| + \frac{2\sqrt{1+\delta_{3T}}}{1-\delta_{3T}} \| \mathbf{e} \| \!\! \right) \nonumber \\
    & \phantom{=} \phantom{=} + \frac{1+\delta_{3T}}{2(1-\delta_{3T})} \| \mbx_{\T_{\text{si}}^{\complement}} \| + \frac{3-\delta_{3T}}{(1-\delta_{3T})^2} \| \mbe \| \nonumber \\
    & \phantom{=} \! \overset{(a)}{\leq} \! \frac{\delta_{3T}(1\!+\!\delta_{3T}\!)^2}{2(1\!-\!\delta_{3T}\!)^4}\! \| \mbx_{\Th_{l-1}^{\complement}} \!\!\| + \frac{1\!+\!\delta_{3T}}{2(1\!-\!\delta_{3T}\!)}\! \| \mbx_{\T_{\text{si}}^{\complement}} \!\| + \frac{4(1\!+\!\delta_{3T}^2\!)}{(1\!-\!\delta_{3T}\!)^3}\! \| \mbe \|. \nonumber
  \end{align}
  In \ensuremath{(a)} we have used for the noise-term that $\frac{(1+\delta_{3T})^2}{2(1-\delta_{3T})^2} \frac{2\sqrt{1+\delta_{3T}}}{1-\delta_{3T}} + \frac{3-\delta_{3T}}{(1-\delta_{3T})^2} \leq \frac{(1+\delta_{3T})^3}{(1-\delta_{3T})^3} + \frac{3-\delta_{3T}}{(1-\delta_{3T})^2} = \frac{4-\delta_{3T} + 4\delta_{3T}^2 + \delta_{3T}^3}{(1-\delta_{3T})^3} \leq \frac{4(1+\delta_{3T}^2)}{(1-\delta_{3T})^3}$. This concludes the proof.
\end{IEEEproof}
\end{proposition}

By using a fixed iteration counter, we use the recurrence inequality in \propositionname~\ref{thrm:recurrence_ipp} to provide the following reconstruction performance bound.

{{
\begin{proposition}[Performance bound of \sipp] \label{thm:perf_bound_pp}
  If $a_{\sipp} < 1$, then after $l^* \to \infty$ iterations, the performance of the \sipp algorithm is bounded by
  \begin{align}
\| \mbx_{\hat{\T}_{\sipp}^{\complement}} \| \leq \frac{b_{\sipp}}{1-a_{\sipp}} \| \mbx_{\T_\text{si}^{\complement}}\| + \frac{c_{\sipp}}{1-a_{\sipp}}  \| \mathbf{e} \|, \label{eqn:thm:perf_bound_ipp1_1}
  \end{align} 
  or
  \begin{align}
    \| \mbx - \hat{\mbx}_{\sipp} \| & \leq \frac{b_{\sipp}}{(1-\delta_{3T})(1-a_{\sipp})} \| \mbx_{\T_\text{si}^{\complement}}\| \nonumber\\ 
      & \phantom{= } + \left( \frac{ c_{\sipp}}{(1-\delta_{3T})(1-a_{\sipp})} +  \frac{1}{\sqrt{1-\delta_{3T}}} \right) \| \mathbf{e} \|. \label{eqn:thm:perf_bound_ipp2_1}
  \end{align} 
  where $a_{\sipp}$, $b_{\sipp}$ and $c_{\sipp}$ are defined in proposition~\ref{thrm:recurrence_ipp}, $\hat{\mbx}_{\sipp}$ and $\hat{\T}_{\sipp}$ are outputs of \algorithmname~\ref{alg:sipp}.
  For a finite $l^*~=~\left\lceil \log\left( \frac{\| \mathbf{e} \|}{\|  \mbx \|}\right)/\log\left( a_{\sipp} \right) \right\rceil $ iterations (with the constraint $\frac{\| \mathbf{e} \|}{\|  \mbx \|} < a_{\sipp}  < 1 $ such that $l^*$ is a positive integer), the performance of the \sipp algorithm is bounded by   
  \begin{align}
\| \mbx_{\hat{\T}_{\sipp}^{\complement}} \| \leq \frac{b_{\sipp}}{1-a_{\sipp}} \| \mbx_{\T_\text{si}^{\complement}}\| + \frac{1 - a_{\sipp} + c_{\sipp}}{1-a_{\sipp}}  \| \mathbf{e} \|, \label{eqn:thm:perf_bound_ipp1}
  \end{align}
  or
  \begin{align}
    \| \mbx - \hat{\mbx}_{\sipp} \| & \leq \frac{b_{\sipp}}{(1-\delta_{3T})(1-a_{\sipp})} \| \mbx_{\T_\text{si}^{\complement}}\| \nonumber\\ 
      & \phantom{= } + \left( \frac{1 - a_{\sipp} + c_{\sipp}}{(1-\delta_{3T})(1-a_{\sipp})} +  \frac{1}{\sqrt{1-\delta_{3T}}} \right) \| \mathbf{e} \|. \label{eqn:thm:perf_bound_ipp2}
  \end{align}  
  \begin{IEEEproof}
    We iteratively apply \propositionname~\ref{thrm:recurrence_ipp} two times:
    \begin{subequations}
      \begin{align}
        \| \mbx_{\hat{\T}_l^{\complement}} \|
        & \leq a_{\sipp} \|\mbx_{\hat{\T}_{l-1}^{\complement}} \|  + b_{\sipp} \| \mbx_{T_\text{si}^{\complement}}\| + c_{\sipp} \| \mathbf{e} \| \nonumber \\
        & \leq a_{\sipp} \left( a_{\sipp} \|\mbx_{\hat{\T}_{l-2}^{\complement}} \|  + b_{\sipp} \| \mbx_{\T_\text{si}^{\complement}}\| + c_{\sipp} \| \mathbf{e} \| \right)  \nonumber \\
        & \phantom{=} + b_{\sipp} \| \mbx_{\T_\text{si}^{\complement}}\| + c_{\sipp} \| \mathbf{e} \| \nonumber \\
        & = a_{\sipp}^{2} \|\mbx_{\hat{\T}_{l-2}^c} \| \! + \! b_{\sipp}\!\!\sum_{i = 0}^{2-1}\!\!{a_{\sipp}^{i}} \| \mbx_{\T_\text{si}^{\complement}}\| \! + \! c_{\sipp}\!\!\sum_{i = 0}^{2-1}\!\!{a_{\sipp}^{i}} \| \mathbf{e} \|. \nonumber 
      \end{align}
    \end{subequations}
      To find a bound on the final performance of \sipp after $l^*$ iterations, we can write
      \begin{subequations}
        \begin{align}
        \|\mbx_{\hat{\T}_{\sipp}^{\complement}}\|
          &  \leq \! a_{\sipp}^{l^*} \|\mbx_{\hat{\T}_{l-l^*}^c} \| \! + \! b_{\sipp}\!\!\sum_{i = 0}^{l^*-1}\!\!{a_{\sipp}^{i}} \| \mbx_{\T_\text{si}^{\complement}}\| \! + \! c_{\sipp}\!\!\sum_{i = 0}^{l^*-1}\!\!{a_{\sipp}^{i}} \| \mathbf{e} \| \nonumber \\
          & \overset{(a)}{\leq} a_{\sipp}^{l^*} \|\mbx \|  + \frac{b_{\sipp}}{1-a_{\sipp}} \| \mbx_{\T_\text{si}^{\complement}}\| + \frac{c_{\sipp}}{1-a_{\sipp}}  \| \mathbf{e} \|.  \nonumber 
        \end{align}
      \end{subequations}
      In \ensuremath{(a)} we have used that $a_{\sipp} < 1$, $\sum_{i = 0}^{l^*-1}\!\!{a_{\sipp}^{i}} \leq \sum_{i = 0}^{\infty}{a_{\sipp}^{i}} = \frac{1}{1-a_{\sipp}}$, and the fact that $\|\mbx_{\hat{\T}_{l-l^*}^{\complement}}\| \leq \|\mbx\|$. Increase in $l^*$ results in exponential decay in $a_{\sipp}^{l^*}$. Letting $l^* \to \infty$, the term $a_{\sipp}^{l^*} \|\mbx \|$ nulls and we get~\eqref{eqn:thm:perf_bound_ipp1_1}. We skip the proof of \eqref{eqn:thm:perf_bound_ipp2_1} due to similarity of the proof of \eqref{eqn:thm:perf_bound_ipp2} shown later. \newline
      Now, for a finite $l^* = \left\lceil \log\left( \frac{\| \mathbf{e} \|}{\|  \mbx \|}\right)/\log\left( a_{\sipp} \right) \right\rceil$, we can write
      \begin{subequations}
        \begin{align} 
        \|\mbx_{\hat{\T}_{\sipp}^{\complement}}\|   
          & {\leq} a_{\sipp}^{l^*} \|\mbx \|  + \frac{b_{\sipp}}{1-a_{\sipp}} \| \mbx_{\T_\text{si}^{\complement}}\| + \frac{c_{\sipp}}{1-a_{\sipp}}  \| \mathbf{e} \|  \nonumber \\
          & {=} a_{\sipp}^{\left\lceil \log\left( \frac{\| \mathbf{e} \|}{\|  \mbx \|}\right)/\log\left( a_{\sipp} \right) \right\rceil} \|\mbx \| \! + \! \frac{b_{\sipp}}{1\!-\!a_{\sipp}} \| \mbx_{\T_\text{si}^{\complement}}\| \! + \! \frac{c_{\sipp}}{1\!-\!a_{\sipp}}  \| \mathbf{e} \| \nonumber \\
          & \leq \frac{\| \mathbf{e} \|}{\| \mbx \|} \| \mbx \| + \frac{b_{\sipp}}{1-a_{\sipp}} \| \mbx_{\T_\text{si}^{\complement}}\| + \frac{c_{\sipp}}{1-a_{\sipp}}  \| \mathbf{e} \|  \nonumber \\
          & = \frac{b_{\sipp}}{1-a_{\sipp}} \|
          \mbx_{\T_\text{si}^{\complement}}\| + \frac{1 - a_{\sipp} +
            c_{\sipp}}{1-a_{\sipp}} \| \mathbf{e} \|. \nonumber
        \end{align}
      \end{subequations}
      Note that as $a_{\sipp} < 1$, we must need the condition $\frac{\| \mathbf{e} \|}{\|  \mbx \|} < a_{\sipp} < 1$ such that $l^* = \left\lceil \log\left( \frac{\| \mathbf{e} \|}{\|  \mbx \|}\right)/\log\left( a_{\sipp} \right) \right\rceil$ is a positive integer.
    The \sipp algorithm uses least squares solution to find $\hat{\mbx}_{\sipp}$. Therefore, to get \eqref{eqn:thm:perf_bound_ipp2} we apply lemma~\ref{lemma:bound_complement} to \eqref{eqn:thm:perf_bound_ipp1}.
    \begin{subequations}
      \begin{align}
        \| \mbx - \hat{\mbx}_{\sipp} \| & \leq \frac{1}{1-\delta_{3T}} \| \mbx_{\hat{\T}_{\sipp}^c} \| + \frac{1}{\sqrt{1-\delta_{3T}}} \|  \mathbf{e} \| \nonumber \\
        & \leq \frac{1}{1\!-\!\delta_{3T}}\!\!\left( \frac{b_{\sipp}}{1\!-\!a_{\sipp}} \| \mbx_{\T_\text{si}^{\complement}}\| + \frac{1\!-\!a_{\sipp}\!+\!c_{\sipp}}{1\!-\!a_{\sipp}}  \| \mathbf{e} \| \right) \nonumber \\
        & \phantom{=} + \frac{1}{\sqrt{1-\delta_{3T}}} \|  \mathbf{e} \| \nonumber \\
        & = \frac{b_{\sipp}}{(1-\delta_{3T})(1-a_{\sipp})} \| \mbx_{\T_\text{si}^{\complement}}\| \nonumber \\
        & \phantom{= } + \left( \frac{1 - a_{\sipp} +
            c_{\sipp}}{(1-\delta_{3T})(1-a_{\sipp})} +
          \frac{1}{\sqrt{1-\delta_{3T}}} \right) \| \mathbf{e} \|. \nonumber
      \end{align}
    \end{subequations}
  \end{IEEEproof}
\end{proposition}
Here we mention that henceforth we will not mention explicit requirements for a finite number of iterations (such as $l^* = \left\lceil \log\left( \frac{\| \mathbf{e} \|}{\|  \mbx \|}\right)/\log\left( a_{\sipp} \right) \right\rceil$) to be a positive integer; necessary requirements can be deciphered from relevant contexts. Using~\eqref{eqn:thm:perf_bound_ipp2_1} we can note that \sipp provides exact reconstruction $\hat{\mbx}_{\sipp} = \mbx$ if $a_{\sipp} < 1$, $\mbx_{\T_\text{si}^{\complement}} = \mathbf{0}$ (that means $\T_\text{si}$ is the true support-set) and $\mathbf{e} = \mathbf{0}$.
}}

{{
\begin{corollary}
  The \sipp algorithm converges to a solution independent of the realization of $\mathbf{x}_{\hat{\mathcal{T}}}$ if and only if
  \begin{align} 
    \delta_{3T} < r , \nonumber
  \end{align}
  where $r = 0.231...$ is a solution to $\frac{r(1+r)^2}{(1-r)^4} = 1$, where $0 < r < 1$.
  \begin{IEEEproof}
    For the \sipp algorithm to converge, it is required that $a_{\sipp} < 1$. We also know that $0 < \delta_{3T} < 1$. We thus solve:
    \begin{subequations}
      \begin{align}
        & & \frac{r(1+r)^2}{(1-r)^4} & = 1 \\
        \Leftrightarrow & & r(1+r)^2 & = (1-r)^4 \nonumber \\
        \Leftrightarrow & & r + 2r^2  + r^3 & = 1 - 4r + 6r^2 - 4r^3 + r^4 \nonumber \\
        \Leftrightarrow & & 0 & = 1 - 5r + 4r^2 -
        5r^3 + r^4. \nonumber 
      \end{align}
    \end{subequations}
    This gives solutions,
    \begin{subequations}
      \begin{align}
        r_{1} = 4.33...,  & ~~~~ r_{2} = 0.219... + i0.976..., \nonumber \\
        r_{3} = 0.231..., & ~~~~ r_{4} = 0.219... - i0.976..., \nonumber 
      \end{align}
    \end{subequations}
    where the only solution $r \triangleq r_{3} = 0.231...$ lies in the interval $0 < \delta_{3T} < 1$. 
    For $a_{\sipp} < 1$, feasible $\delta_{3T} < r$.
  \end{IEEEproof}
\end{corollary}
\begin{example} \label{example:sipp1}
  In this example 
, if we have $\delta_{3T} = 0.17$, we get $a_{\sipp} < 0.50$, $b_{\sipp} < 0.71$, $c_{\sipp} < 7.20$. Then according to proposition~\ref{thm:perf_bound_pp}, after $l^* = \left\lceil \log\left( \frac{\| \mathbf{e} \|}{\| \mbx \|}\right)/\log\left( 0.5 \right) \right\rceil$ iterations the performance of \sipp fulfills the following bounds
  \begin{align}
    \| \mbx_{\hat{\T}_{\sipp}^{\complement}} \| & < 1.42 \| \mbx_{\T_\text{si}^{\complement}}\| + 15.2 \| \mathbf{e} \|, \nonumber
  \end{align}
or
  \begin{align}
    \| \mbx - \hat{\mbx}_{\sipp} \| & < 1.72 \| \mbx_{\T_\text{si}^{\complement}}\| + 19.4 \| \mathbf{e} \|. \nonumber
  \end{align}
  By this example we see that \sipp reaches its performance bound when $\delta_{3T} \leq 0.17$ whereas \sp requires $\delta_{3T} \leq 0.139$~\cite{Giryes:rip_based_near_oracle_trans} for a similar result.
\end{example}
\begin{example} \label{example:sipp2}
  If instead $\delta_{3T} = 0.23$, then after $l^* = \left\lceil \log\left( \frac{\| \mathbf{e} \|}{\| \mbx \|}\right)/\log\left( 0.99 \right) \right\rceil$ iterations the performance of \sipp fulfills the following bounds
  \begin{align}
    \| \mbx_{\hat{\T}_{\sipp}^{\complement}} \| & < 78.8 \| \mbx_{\T_\text{si}^{\complement}}\| + 912 \| \mathbf{e} \|, \nonumber
  \end{align}
or
  \begin{align}
    \| \mbx - \hat{\mbx}_{\sipp} \| & < 95.4 \| \mbx_{\T_\text{si}^{\complement}}\| + 1.19 \cdot 10^3 \| \mathbf{e} \|. \nonumber
  \end{align}
  Note that $\delta_{3T} = 0.23$ is close to $r = 0.231...$, and a theoretical convergence is not guaranteed for $\delta_{3T} \geq r = 0.231...$.
\end{example}
}}

\subsection{Analysis of fusion} \label{sec:consensus}
Fusion has two parts: \consensus and \expansion. The strategy of \consensus is based on a voting principle, which in general is non-trivial to analyze due to the counting of non-negative integers followed by decision. The \consensus endeavors to estimate the joint support part for sensor node $\p$ as $\Jh_{\p}$. Following \algorithmname~\ref{alg:fusion}, we note that
\begin{align}
  \Jh_{\p} = \left\{ i: i\in\left( (\Thi \cap \Thii) \cup (\Thii \cap \Thiii) \right), \,\, \forall \q,\r \in \mathcal{L}_{\p}^{\text{in}}, \q \neq \r \right\} \nonumber
\end{align}
Any index $i\in\Ti$ is referred to as a correct index for node $\p$. 
{{
At this point, we use a notion of probability for ease of understanding and arguments, and not in a rigorous sense. 
}}
Let us denote the probability of an index from the output of the \sipp algorithm $i \in \Thi$ to be correct by the notation $\P{i\in\Ti | i\in\Thi}$, and the probability of index $i \in \Jh_{\p}$ to be correct by the notation $\P{i\in\Ti | i\in\Jh_{\p}}$. Following the voting strategy in \consensus of \algorithmname~\ref{alg:fusion}, we introduce the following assumption.
\begin{assumption}
\label{assump:prob_comm_supp}
$\Jh_{\p}$ is at-least as reliable as $\Thi$ in a probabilistic sense. That is
\begin{eqnarray}
 \P{i\in\Ti | i\in\Jh_{\p}} \geq \P{i\in\Ti|i\in\Thi}. \nonumber
\end{eqnarray}
\wqed 
\end{assumption}
{{
A rigorous proof on the validity of \assumptionname~\ref{assump:prob_comm_supp} is recently addressed by us in \cite{Sundman:democratic_vote} (see propositions 4 and 5, and remark 1 of \cite{Sundman:democratic_vote}).
The \assumptionname~\ref{assump:prob_comm_supp} motivates the inclusion of $\Jh_{\p}$ in $\T_{{\p},\text{si}}$ which is the output of \expansion in \algorithmname~\ref{alg:fusion}. Note that \expansion provides an estimate of $\Ti$ as $\T_{{\p},\text{si}} = \hat{\I}_{\p} \cup \hat{\J}_{\p}$. To maintain the cardinality of $|\T_{{\p},\text{si}}| = T$, the \expansion algorithm discards $\Thi \setminus \hat{\I}_{\p}$. That means, in our design of \algorithmname~\ref{alg:fusion} we have more trust in the signal coefficients associated with $\Jh_{\p}$ than the signal coefficients associated with $\Thi \setminus \hat{\I}_{\p}$.
Therefore, in pursuit of further analytical progress we introduce the following assumption.
}}
\begin{assumption}
\label{assumption:expansion}
 Signal coefficients associated with $\Jh_{\p}$ contains at-least as much energy than the signal coefficients associated with the discarded $\Thi \setminus \hat{\I}_{\p}$. That is
 \begin{align}
    \| (\mbx_{\p})_{\Jh_{\p}} \|^2 \geq \| (\mbx_{\p})_{\Thi \setminus \hat{\I}_{\p}} \|^2. \nonumber 
\end{align} \wqed
\end{assumption}
{{
In practice we verified by a simulation experiment that \assumptionname~\ref{assumption:expansion} holds for most of realizations $\mbx_{\p}$, but not for all realizations (the simulation experiment is not reported in the paper). Compliance of \assumptionname~\ref{assumption:expansion} for a realization is a sufficient condition in our worst case analysis approach.
}}

Using \assumptionname~\ref{assumption:expansion}, we will in \propositionname~\ref{thm:merger} characterize the performance bound of the \expansion algorithm. First, we introduce the following lemma.
\begin{lemma}\label{lemma:merge}
  \begin{align}
        \| (\mbx_{\p})_{\T_{\p,\text{si}}^{\complement}} \|^2 = \| (\mbx_{\p})_{\hat{\T}_{\p}^{\complement}}\|^2 + \| (\mbx_{\p})_{\hat{\T}_{\p} \setminus \hat{\I}_{\p}} \|^2 - \| (\mbx_{\p})_{\hat{\J}_{\p}} \|^2 \nonumber 
  \end{align}
\textit{Proof:} In the proof, we drop notation $\p$.
We have that $\hat{\I} \subseteq \hat{\T}$ and $\hat{\I} \subseteq \T_{\text{si}}$ by construction and definition, respectively. For the initial support-set we have that:
  \begin{align}
    \| \mbx_{\T_{\text{si}}^{\complement}} \|^2 & = \| \mbx_{(\hat{\I} \cup \hat{\J})^{\complement}} \|^2 \nonumber \\
    & \overset{(a)}{=} \| \mbx_{\hat{\I}^{\complement} \cap \hat{\J}^{\complement}} \|^2 \nonumber \\
    & = \| \mbx_{\hat{\I}^{\complement} \setminus \hat{\J}} \|^2 \nonumber \\
    & \overset{(b)}{=} \| \mbx_{\hat{\I}^{\complement}} \|^2 - \| \mbx_{\hat{\J}} \|^2. \label{eqn:merge1}
  \end{align}
In \ensuremath{(a)} we have used De Morgan's law and in \ensuremath{(b)} we have used that $\Jh \subset \hat{\I}^{\complement}$. By the same trick we also have that
\begin{align}
  \| \mbx_{\hat{\T}^{\complement}} \|^2 = \| \mbx_{\hat{\I}^{\complement}} \|^2 - \| \mbx_{\hat{\T}\setminus\hat{\I}} \|^2 \label{eqn:merge2}
\end{align}
Putting \eqref{eqn:merge2} into \eqref{eqn:merge1}, we get
  \begin{align}
    \| \mbx_{\T_{\text{si}}^{\complement}} \|^2 & = \| \mbx_{\hat{\I}^{\complement}} \|^2 - \| \mbx_{\hat{\J}} \|^2 \nonumber \\
    & = \| \mbx_{\hat{\T}^{\complement}} \|^2 + \| \mbx_{\hat{\T}\setminus\hat{\I}} \|^2 - \| \mbx_{\hat{\J}} \|^2. \nonumber
  \end{align}
\qed
\end{lemma}

\begin{proposition}[Performance bound for expansion] \label{thm:merger}
  \begin{align}
    \| (\mbx_{\p})_{\T_{{\p},\text{si}}^{\complement}} \| \leq a_{\texttt{co}} \| (\mbx_{\p})_{\hat{\T}_{\p}^{\complement}} \|, \nonumber
  \end{align}
where
\begin{align}
  a_{\texttt{co}} \leq 1. \nonumber
\end{align}
\begin{IEEEproof}  In the proof, we drop notation $\p$.
  \begin{align}
    \| \mbx_{\T_{\text{si}}^{\complement}} \| & \overset{(a)}{=} \sqrt{\| \mbx_{\hat{\T}^{\complement}} \|^2 + \| \mbx_{\hat{\T}\setminus\hat{\I}} \|^2 - \| \mbx_{\hat{\J}} \|^2} \nonumber \\
    & \overset{(b)}{\leq} \sqrt{\| \mbx_{\hat{\T}^{\complement}} \|^2} = \| \mbx_{\hat{\T}^{\complement}} \| \label{eqn:alpha_define}
  \end{align}
  In \ensuremath{(a)}, we applied \lemmaname~\ref{lemma:merge} and in \ensuremath{(b)}, we used \assumptionname~\ref{assumption:expansion}. Based on this argument, we introduce the constant $a_{\texttt{co}} \leq 1$ such that
  \begin{align}
        \| \mbx_{\T_{\text{si}}^{\complement}} \| & \leq a_{\texttt{co}} \| \mbx_{\hat{\T}^{\complement}} \| \leq \| \mbx_{\hat{\T}^{\complement}} \|.
  \end{align}
  When the quality of information exchange over network is good, meaning that the inequality in \eqref{eqn:alpha_define} is large, then $a_{\texttt{co}}$ is small and vice versa.
\end{IEEEproof}
\end{proposition}

{{
\begin{remark}
 The parameter $a_{\texttt{co}} \leq 1$ characterizes the quality of fusion (combined effect of \consensus and \expansion), that means the quality of information exchange. Quality of information exchange directly depends on network connectivity. A low value of $a_{\texttt{co}}$ corresponds to a good quality of information exchange.
\end{remark}
}}

\subsection{Distributed Parallel Pursuit}
Now we will characterize a performance bound for \dipp. In order to do so, we first derive the recurrence inequality of \dipp by using the performance bounds of \sipp and \consensus algorithms.

{{
\begin{proposition}[Recurrence inequality of \dipp] \label{prop:recurrence_dipp} When $a_{\sipp} < 1$, the recurrence inequality of \dipp is
  \begin{align}
  \| \mbx_{\hat{\T}_{\p,k}^{\complement}} \| < \frac{a_{\texttt{co}}b_{\sipp}}{1-a_{\sipp}} \| \mbx_{\hat{\T}_{\p,k-1}^{\complement}} \| + \frac{1 - a_{\sipp} + c_{\sipp}}{1-a_{\sipp}}  \| \mbe_{\p} \|. \nonumber
  \end{align}
Here $a_{\sipp}$, $b_{\sipp}$, $c_{\sipp}$ are parameters associated with the underlying \sipp algorithm and the parameters are defined in Proposition~\ref{thrm:recurrence_ipp}. The parameter $a_{\texttt{co}}$ is defined in Proposition~\ref{thm:merger}.
Note that 
\begin{align}
  {a_{\texttt{co}}} < 1, \nonumber
\end{align}
and $a_{\sipp}$, $b_{\sipp}$, $c_{\sipp}$ are functions of $\delta_{3T}$ as
\begin{align}
  a_{\sipp} & \triangleq \frac{\delta_{3T} (1+\delta_{3T})^2}{(1-\delta_{3T})^4} > 0, \nonumber \\
  b_{\sipp} & \triangleq \frac{1 + \delta_{3T}}{2(1-\delta_{3T})} > 0, \nonumber \\
  c_{\sipp} & \triangleq \frac{4(1+\delta_{3T})}{(1-\delta_{3T})^3} > 0. \nonumber 
\end{align}
\begin{IEEEproof}
  Proof in \appendixname~\ref{app:dipp}.
\end{IEEEproof}
\end{proposition}
}}

{{
Using the recurrence inequality, we can now derive the performance bound of \dipp.
\begin{proposition}[Performance bound of \dipp]\label{prop:performance_dipp}
  If $a_{\sipp} < 1$ and ${a_{\texttt{co}}} \frac{b_{\sipp}}{1-a_{\sipp}} < 1$, then after $k^*~=~\left\lceil \log\left( \frac{\| \mbe_{\p} \|}{\|  \mbx_{\p} \|}\right)/\log\left( {a_{\texttt{co}}} \frac{b_{\sipp}}{1-a_{\sipp}} \right) \right\rceil$ iterations (with the constraint that $\frac{\| \mbe_{\p} \|}{\|  \mbx_{\p} \|} < {a_{\texttt{co}}} \frac{b_{\sipp}}{1-a_{\sipp}} < 1$ such that $k^*$ is a positive integer), the performance of the \dipp algorithm is bounded by:
  \begin{align}
    \| (\mbx_{\p})_{\hat{\T}_{\p,\dipp}^{\complement}} \| \leq \left( 1 + \frac{1-a_{\sipp} + c_{\sipp}}{1-a_{\sipp}-{a_{\texttt{co}}}b_{\sipp}} \right) \| \mbe_{\p} \| \label{dipp:perf1} 
  \end{align}
  or
  \begin{align}
    \|  \mbx_{\p} - \hat{\mbx}_{\p,\dipp} \| \leq & \left( \frac{1-a_{\sipp} + c_{\sipp}}{(1-\delta_{3T})(1-a_{\sipp}-{a_{\texttt{co}}}b_{\sipp})} \right. \nonumber \\
    & \phantom{=} \left. + \frac{2}{1-\delta_{3T}} \right) \| \mbe_{\p} \| \label{dipp:perf2}
  \end{align}
  where the constants are defined in Proposition~\ref{prop:recurrence_dipp}. 
  Note that $a_{\sipp}$, $b_{\sipp}$, $c_{\sipp}$ are function of $\delta_{3T}$, 
  and hence the upper bounds are functions of $\delta_{3T}$, $a_{\texttt{co}}$ and $\mbe_{\p}$.
  \begin{IEEEproof}
    Proof in \appendixname~\ref{app:dipp}.
  \end{IEEEproof}
\end{proposition}
}}

\begin{example} Using $\delta_{3T} = 0.17$ as in Example~\ref{example:sipp1} and $a_{\texttt{co}} = 0.27$ gives that $a_{\sipp} < 0.5$ and ${a_{\texttt{co}}} \frac{b_{\sipp}}{1-a_{\sipp}} < 0.5$. This translates into the following bounds
  \begin{align}
    \| (\mbx_{\p})_{\hat{\T}_{\p,\dipp}^{\complement}} \| < 28.3 \|  \mbe_{\p} \|, \nonumber
  \end{align}
  or,
  \begin{align}
    \|  \mbx_{\p} - \hat{\mbx}_{\p,\dipp} \| < 36.5 \|  \mbe_{\p} \|. \nonumber
  \end{align}
\end{example}

\begin{example}
  Using $\delta_{3T} = 0.23$ as in Example~\ref{example:sipp2} and $a_{\texttt{co}} = 1.61\cdot10^{-4}$ gives that $a_{\sipp} < 0.99$ and ${a_{\texttt{co}}} \frac{b_{\sipp}}{1-a_{\sipp}} < 1$. This translates into the following bounds
  \begin{align}
    \| (\mbx_{\p})_{\hat{\T}_{\p,\dipp}^{\complement}} \| < 1.08 \cdot 10^{3} \| \mbe_{\p} \|, \nonumber
  \end{align}
  or,
  \begin{align}
    \|  \mbx_{\p} - \hat{\mbx}_{\p,\dipp} \| < 1.41 \cdot 10^{3} \| \mbe_{\p} \|. \nonumber
  \end{align}
\end{example}

{{
In the \dipp algorithm, there are two parameters that impact the performance: the classical RIC $\delta_{3T}$ of measurement matrix, and the (new) $a_{\texttt{co}}$ that characterizes performance of quality of fusion (quality of information exchange by \consensus and \expansion). If both parameters are good (that means small $\delta_{3T}$ and $a_{\texttt{co}}$), then the \dipp algorithm will perform good. One main advantage of the \dipp algorithm compared to a standard (disconnected) algorithm is that we can allow $\delta_{3T}$ to be higher than in a single-node case while still providing a performance bound independent of signal realization, provided that $a_{\texttt{co}}$ is smaller. Note that a high $\delta_{3T}$ results in a poor measurement quality, but its debilitating effect on reconstruction performance is compensated by a good quality of information exchange.
}}

\section{Simulation Results} \label{sec:numerical}
{{
We have provided analytical performance bounds for \dipp and the underlying sub-algorithms. These bounds put worst case restrictions on the system in terms of RIC ($\delta_{3T}$) and quality of information exchange ($a_{\texttt{co}}$). In practice, it turns out that reconstruction algorithms for \cs usually perform well at significantly less restrictive set-ups. We already mentioned that quality of information exchange improves with increase in network connectivity. Therefore our hypothesis is: performance of \dipp improves with increase in network connectivity. In this section we mainly verify the hypothesis. 
}}

\begin{figure*}[t]
  \centering
  \subfloat[\asce vs fraction of measurements]{
    \resizebox{1\columnwidth}{!}{
      \includegraphics[width=\columnwidth]{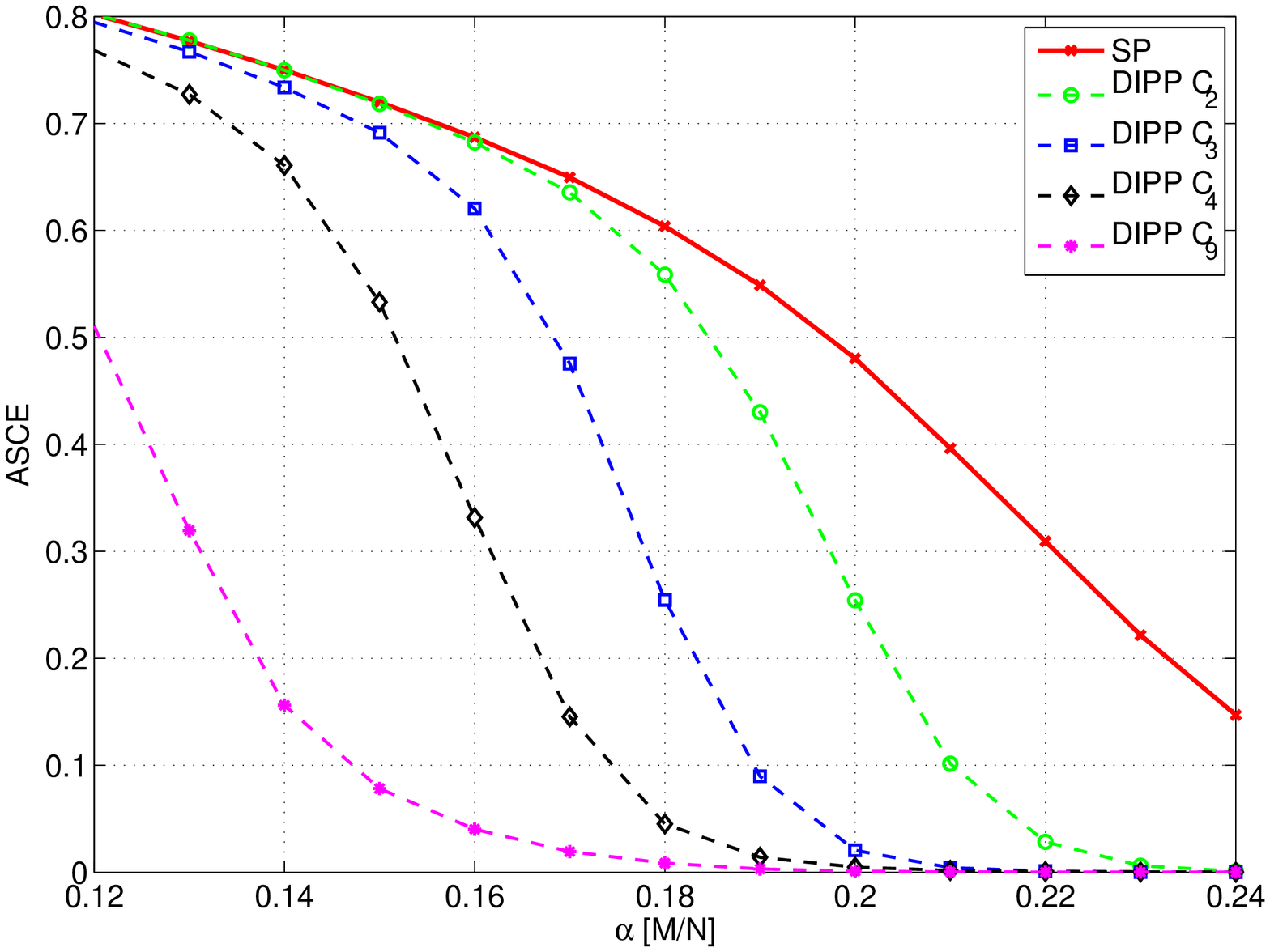} 
      \label{fig:noisy_binary_asce}
    }
  }
  \subfloat[\srer vs fraction of measurements]{
    \resizebox{1\columnwidth}{!}{
      \includegraphics[width=\columnwidth]{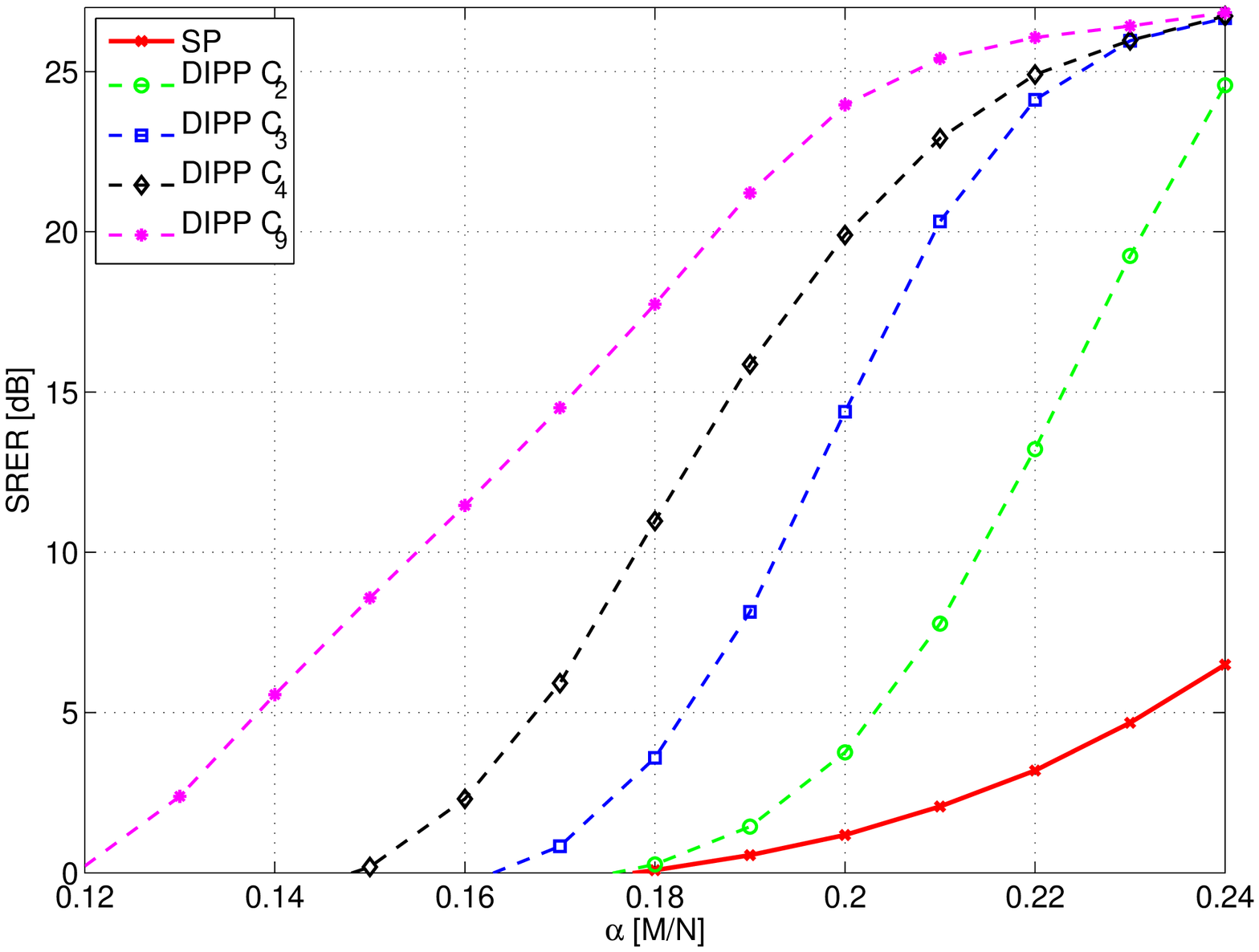}
      \label{fig:noisy_binary_srer}
    }
  }
  \caption{Reconstruction of binary sparse signals at $\smnr = 20$ dB, by varying network connectivity and fraction of measurements $\alpha$.}
  \label{fig:noisy_binary}
\end{figure*}

\subsection{Performance Measures and Experimental Setups}
We use two performance measures. The first performance measure is called signal-to-reconstruction-error ratio (\srer), defined as
\begin{eqnarray}
\srer = \frac{ \mathcal{E} \{ \| \mathbf{x} \|_{2}^{2} \} }{\mathcal{E} \{ \| \mathbf{x} - \hat{\mathbf{x}} \|_{2}^{2} \} },  \nonumber
\end{eqnarray} 
where $\mathcal{E}$ is the expectation taken over all nodes and all realizations. Our objective is to achieve a higher \srer. Note that the \srer is the inverse of normalized mean square error.

We also define a measure which provides a direct evaluation of the support-set recovery performance. This is a distortion measure $d(\T,\hat{\T})= 1- \left( | \T \cap \hat{\T} | / |\T|  \right)$ \cite{Gastpar:a_note_on_optimal_support_recovery}, which we recently used in \cite{Chatterjee:projection_based_look_ahead}. Here, $\T_{\p}$ is the local support-set, that is $\T_{\p} = \J \cup \I_{\p}$. Considering a large number of realizations, we can compute the average of $d(\T_{\p},\hat{\T}_{\p})$. Based on this distortion, we define the average support-set cardinality error (\asce) as follows
\begin{align}
\asce = \mathcal{E}\left\{ d(\T, \hat{\T}) \right\}= 1 - \mathcal{E}\left\{\frac{|\T \cap \hat{\T}|}{|\T|} \right\}. \nonumber
\end{align}
Note that the \asce has the range $[0, 1]$ and our objective is to achieve a lower \asce. Along-with \srer, the \asce is used as the second performance evaluation measure because the principle of greedy algorithms is to estimate the underlying support-set. We perform average based empirical testing, where \srer and \asce are computed for large sets of data. To measure the level of under-sampling, we define the fraction of measurements
\begin{align}
\alpha = \frac{M}{N}. \nonumber
\end{align}
For a given network topology $\C_{i}, \,\, i \in [0,|\mathcal{L}|-1]$ (see \sectionname~\ref{sec:network_model}), the steps of testing strategy are listed as follows:
\begin{packed_enum}
\item Given the parameters $N$, $T$ choose an $\alpha$ (such that $M$ is an integer). \label{item:L}
\item Randomly generate a set of $M \times N$ sensing matrices $\left\{\mathbf{A}_{\p}\right\}_{\p=1}^{|\mathcal{L}|}$ where the components are drawn independently from an i.i.d. Gaussian source (i.e. $a_{k,l} \sim \mathcal{N}\left( 0, \frac{1}{M} \right)$) and then scale the columns of $\mathbf{A}_{\p}$ to unit-norm. \label{item:sensing_matrix}
\item Randomly generate a set of signal vectors $\{\mathbf{x}_{\p}\}_{l=p}^{|\mathcal{L}|}$ following Section~\ref{sec:mixed_support_signal_model}. The common and private support-sets are chosen uniformly over the set $\{1,2, \ldots, N \}$. The non-zero components of $\mathbf{x}_{\p}$ are independently drawn by either of the following two methods. 
\begin{enumerate}
\item The non-zero components are drawn independently from a standard Gaussian source. This type of signal is referred to as Gaussian sparse signal.
\item The non-zero components are set to ones. This type of signal is referred to as binary sparse signal.
\end{enumerate}
Note that the Gaussian sparse signal is compressible in nature, meaning that, in the descending order, the sorted amplitudes of a Gaussian sparse signal vector's components decay fast with respect to the sorted indices. This decaying trend corroborates with several natural signals (for example, wavelet coefficients of an image). On the other hand, a binary sparse signal is not compressible in nature, but of special interest for comparative study, since it represents a particularly challenging case for greedy reconstruction strategies \cite{Tropp:signal_recovery}, \cite{Dai:subspace_pursuit}.  \label{item:data}
\item Compute the measurements $\mathbf{y}_{\p} = \mathbf{A}_{\p}\mathbf{x}_{\p} + \mathbf{e}_{\p}, \forall \p \in \mathcal{L}$. Here $\mathbf{e}_{\p} \sim \mathcal{N}(\mathbf{0},\sigma_{e}^2 \mathbf{I}_M)$.
\item Apply the \cs algorithms on the data $\{\mathbf{y}_{\p}\}_{\p=1}^{|\mathcal{L}|}$ independently. 
\end{packed_enum}
In the above simulation procedure, for each node $\p \in \mathcal{L}$, $10^2$ realizations of sensing matrices were used, and, for each sensing node, $10^2$ realizations of data vectors were used. We used $10$ nodes in the network. Thus, the performance is averaged over $10 \cdot 100 \cdot 100 = 10^5$ data.

Considering the measurement noise $\mathbf{e}_{\p} \sim \mathcal{N} \left( \mathbf{0}, \sigma_{e}^{2} \mathbf{I}_{M} \right)$, we define the signal-to-measurement-noise-ratio (\smnr) as
\begin{eqnarray}
\smnr = \frac{ \mathcal{E} \{ \| \mathbf{x}_{\p} \|_{2}^{2} \} }{ \mathcal{E} \{ \| \mathbf{e}_{\p} \|_{2}^{2} \} } , \nonumber
\end{eqnarray} 
where $\mathcal{E} \{ \| \mathbf{e}_{\p} \|_{2}^{2} \} = \sigma_{e}^{2}M$. For noisy measurement case, we report the experimental results at \smnr $20$ dB. 

In the convergence and performance results we have used the signal dimensionality $N = 1000$, $J = 15$ and $I_{\p} = 5$, giving $T = 20$. Such a $2\%$ sparsity level is chosen in accordance with real life scenarios, for example most of the energy of an image signal in the wavelet domain is concentrated within $2-4\%$ coefficients \cite{Candes:an_introduction}.

\begin{figure*}[t]
  \centering
  \subfloat[\asce vs fraction of measurements]{
    \resizebox{1\columnwidth}{!}{
      \includegraphics[width=\columnwidth]{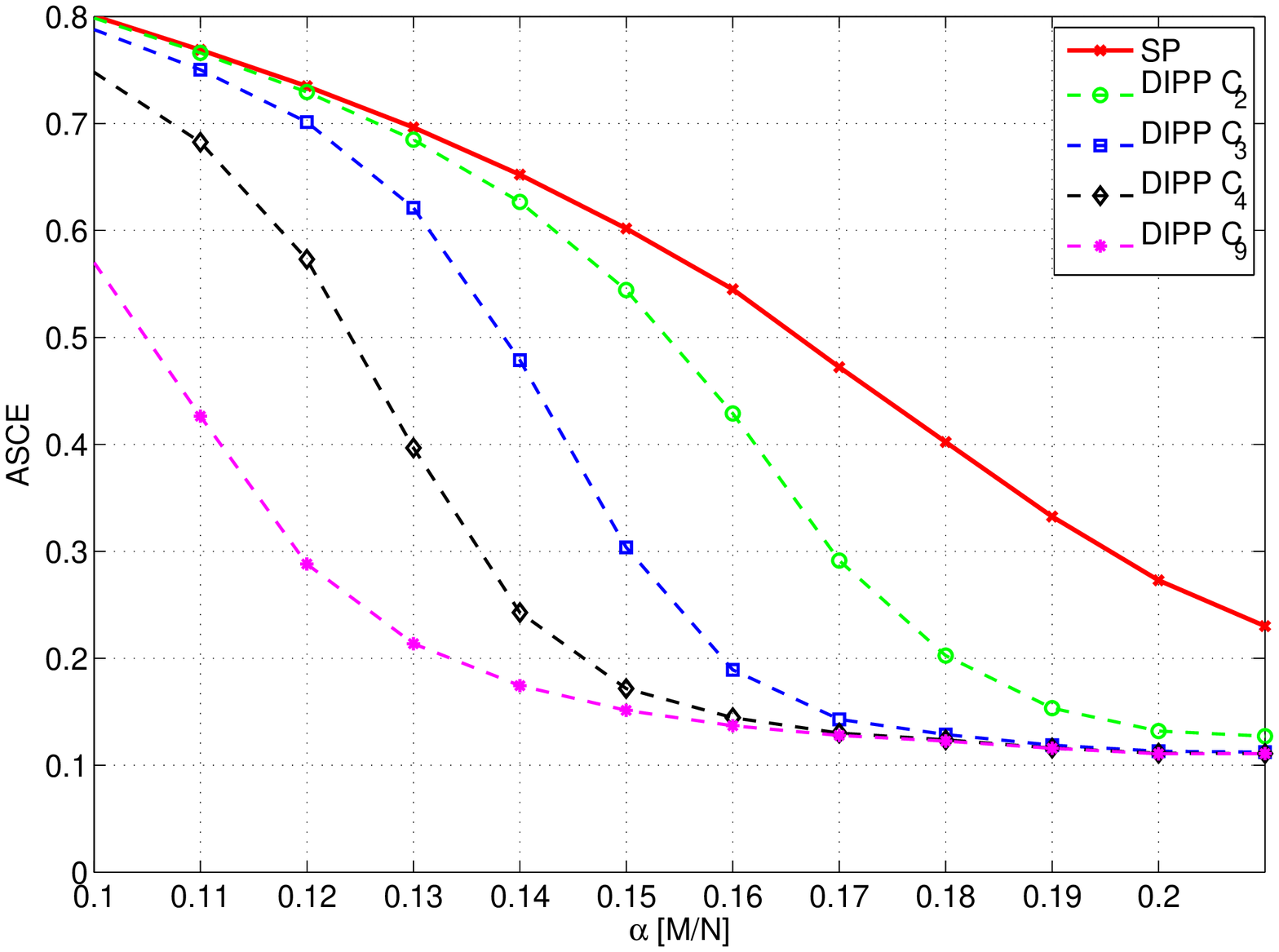}
      \label{fig:noisy_gaussian_asce}
    }
  }
  \subfloat[\srer vs fraction of measurements]{
    \resizebox{1\columnwidth}{!}{
      \includegraphics[width=\columnwidth]{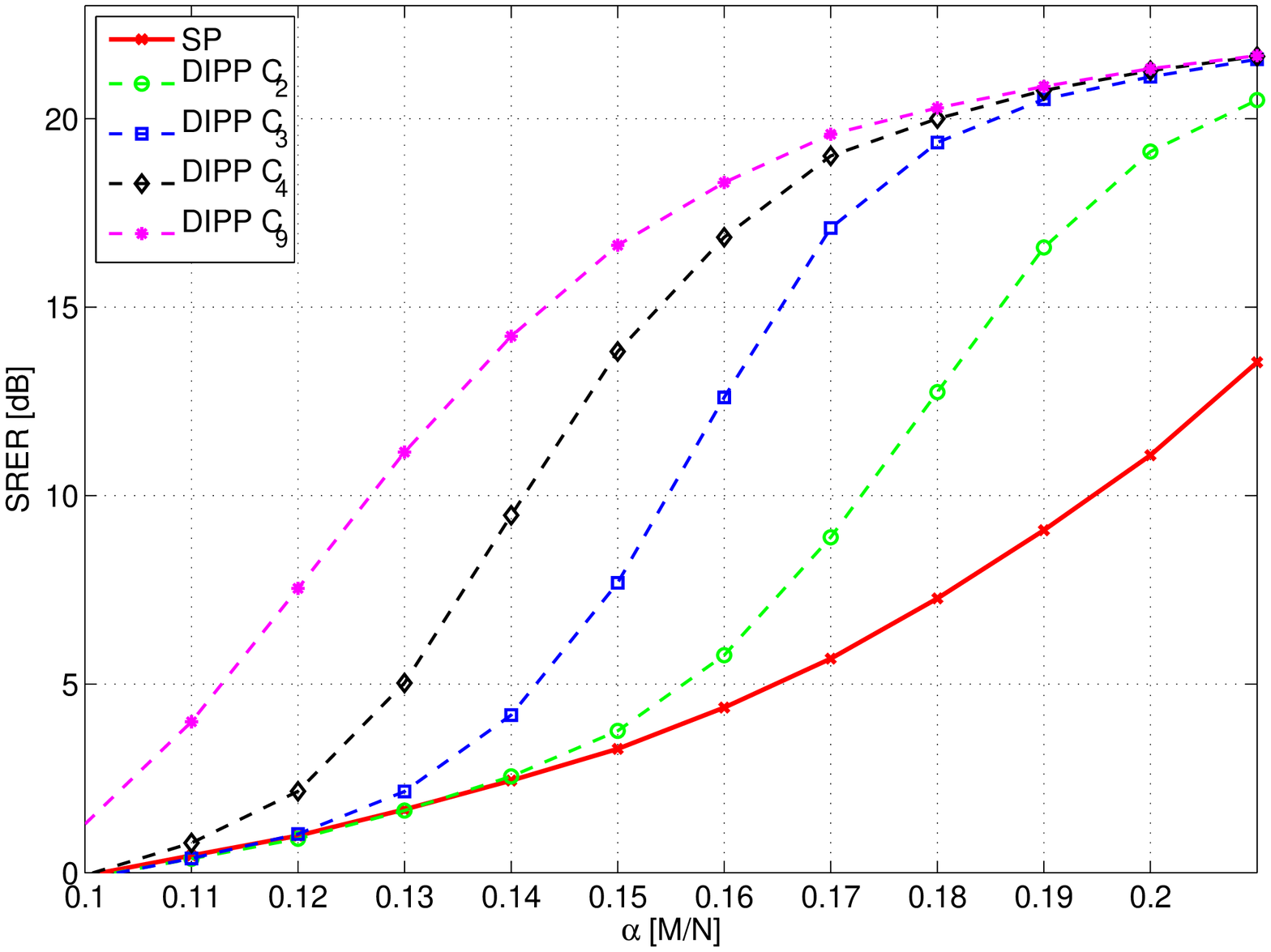}
      \label{fig:noisy_gaussian_srer}
    }
  }
  \caption{Reconstruction of Gaussian sparse signals at $\smnr = 20$ dB, by varying network connectivity and fraction of measurements $\alpha$.
  }
  \label{fig:noisy_gaussian}
\end{figure*}

\begin{figure*}[t]
  \centering
  \subfloat[\asce vs fraction of measurements]{
    \resizebox{1\columnwidth}{!}{
      \includegraphics[width=\columnwidth]{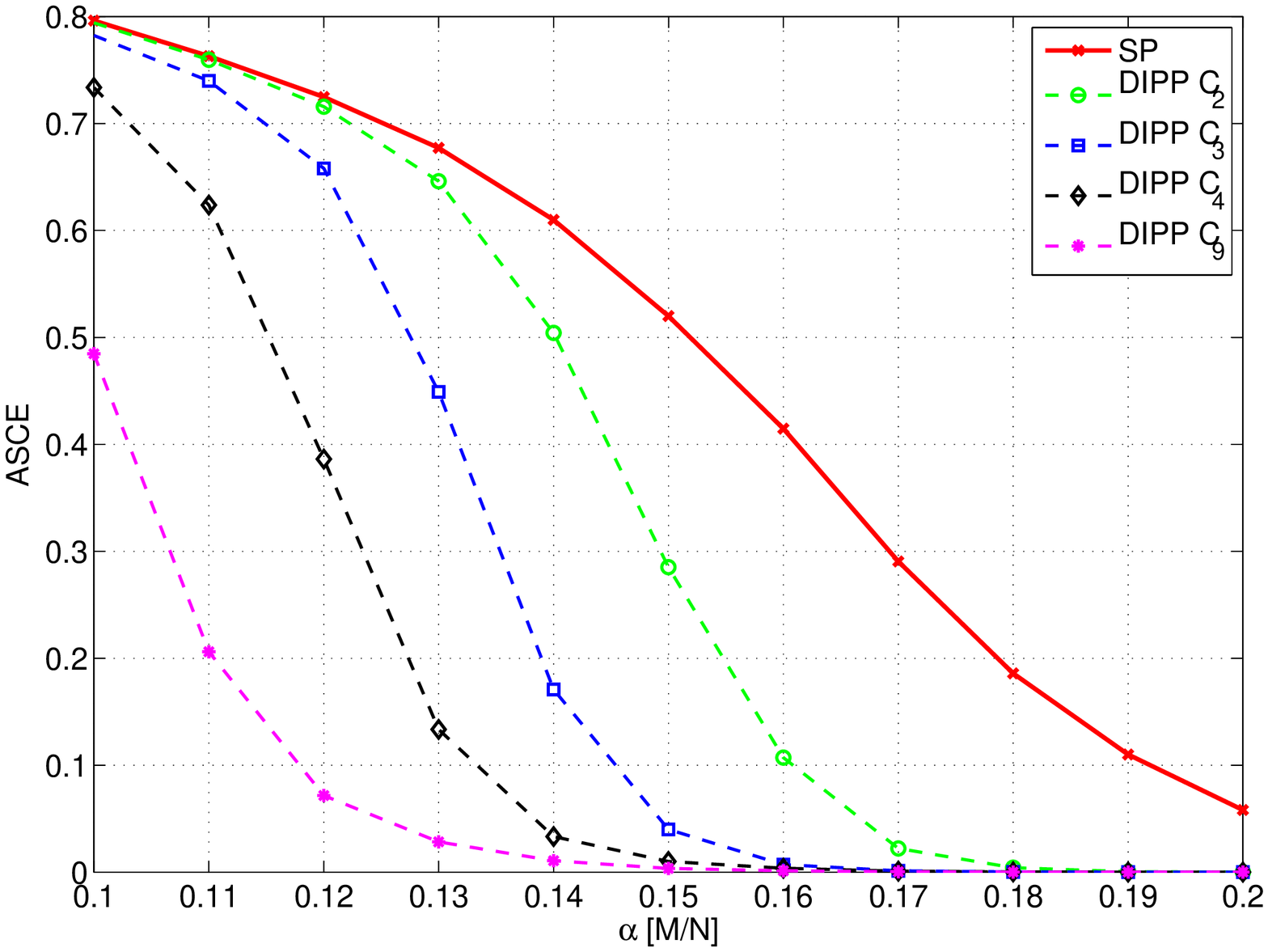}
      \label{fig:clean_gaussian_asce}
    }
  }
  \subfloat[\srer vs fraction of measurements]{
    \resizebox{1\columnwidth}{!}{
  \includegraphics[width=\columnwidth]{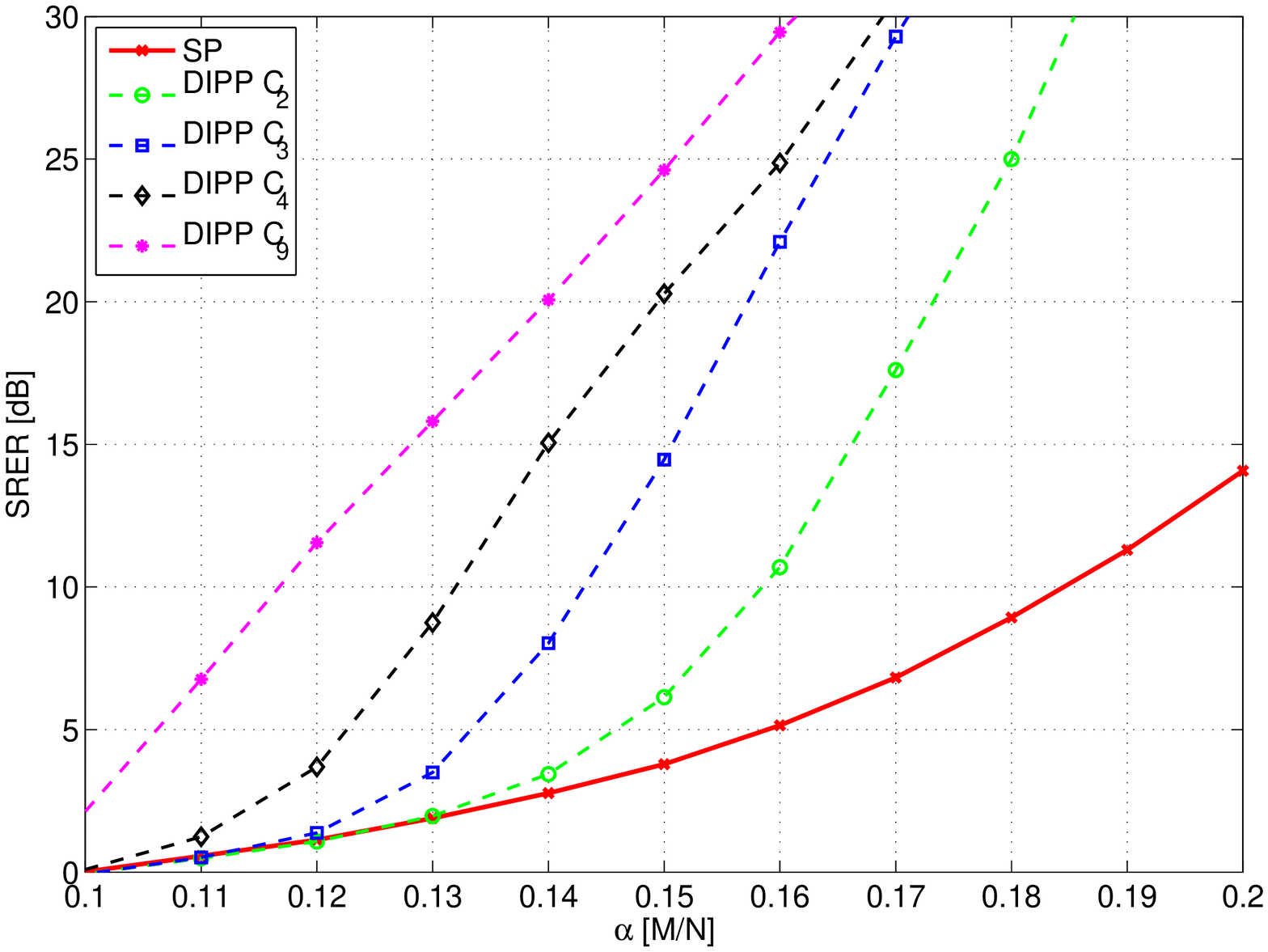}
      \label{fig:clean_gaussian_srer}
    }
  }
  \caption{Reconstruction of Gaussian sparse signals at clean condition, by varying network connectivity and fraction of measurements $\alpha$. 
  }
  \label{fig:clean_gaussian}
\end{figure*}

\subsection{Performance Results}
We now provide the average performance results using the performance measures \srer and \asce described earlier. We use a non-decreasing norm stopping criterion in algorithms. 
{{
We report the results in five parts. We used the structured network model of \sectionname~\ref{sec:network_model} for the first four parts and the Watts-Strogatz~\cite{watts1998collective} network model for the fifth part. The structured network used 10 nodes and the Watts-Strogatz network used 100 nodes. The performance of \sp is included in all experiments as a benchmark characterizing a single-sensor (disconnected) scenario.
}}

\subsubsection{Binary signals with additive noise}
In \figurename~\ref{fig:noisy_binary} we provide performance results for \dipp using binary sparse signals with $\smnr = 20$ dB. We note that the performance of the system improves significantly as the connectivity in the network improves (remember that $a_{\texttt{co}}$ is the network parameter). Similarly, we see improvement in the system with growing $\alpha$ because a growing $\alpha$ improves the RIC $\delta_T$ parameter. Observe that \asce tends to zero as $\alpha$ increases which means that there will in average be no support-set errors after some point.

\subsubsection{Gaussian signals with additive noise}
In \figurename~\ref{fig:noisy_gaussian} we present the results for Gaussian sparse signals under $\smnr = 20$ dB. Here also we notice significant performance improvement. For example, in \figurename~\ref{fig:noisy_gaussian_srer}, at $\alpha = 0.16$, \dipp in a network of $4$ neighbors provides almost $13$ dB performance gain over \sp. As we expect, the \asce never reaches $0$ since in a local node, strong noise may be mistaken for signal components.

\subsubsection{Gaussian signals in clean condition}
\figurename~\ref{fig:clean_gaussian} shows performance for Gaussian sparse signals in clean environment (no measurement noise). Here, we are expected to achieve perfect signal recovery for better system conditions, such as a higher $\alpha$ and/or a higher network connectivity. Therefore, the \srer may theoretically reach $\infty$ which in our case will be at the level of machine precision. In such case, the \asce may provide a better reference curve since it tends to zero rather than $\infty$. We found that such perfect reconstruction quality can be achieved by increasing network connectivity at a fixed $\alpha$. 

\begin{figure*}[ht]
  \centering
  \subfloat[\asce vs \smnr]{
    \resizebox{1\columnwidth}{!}{
      \includegraphics[width=\columnwidth]{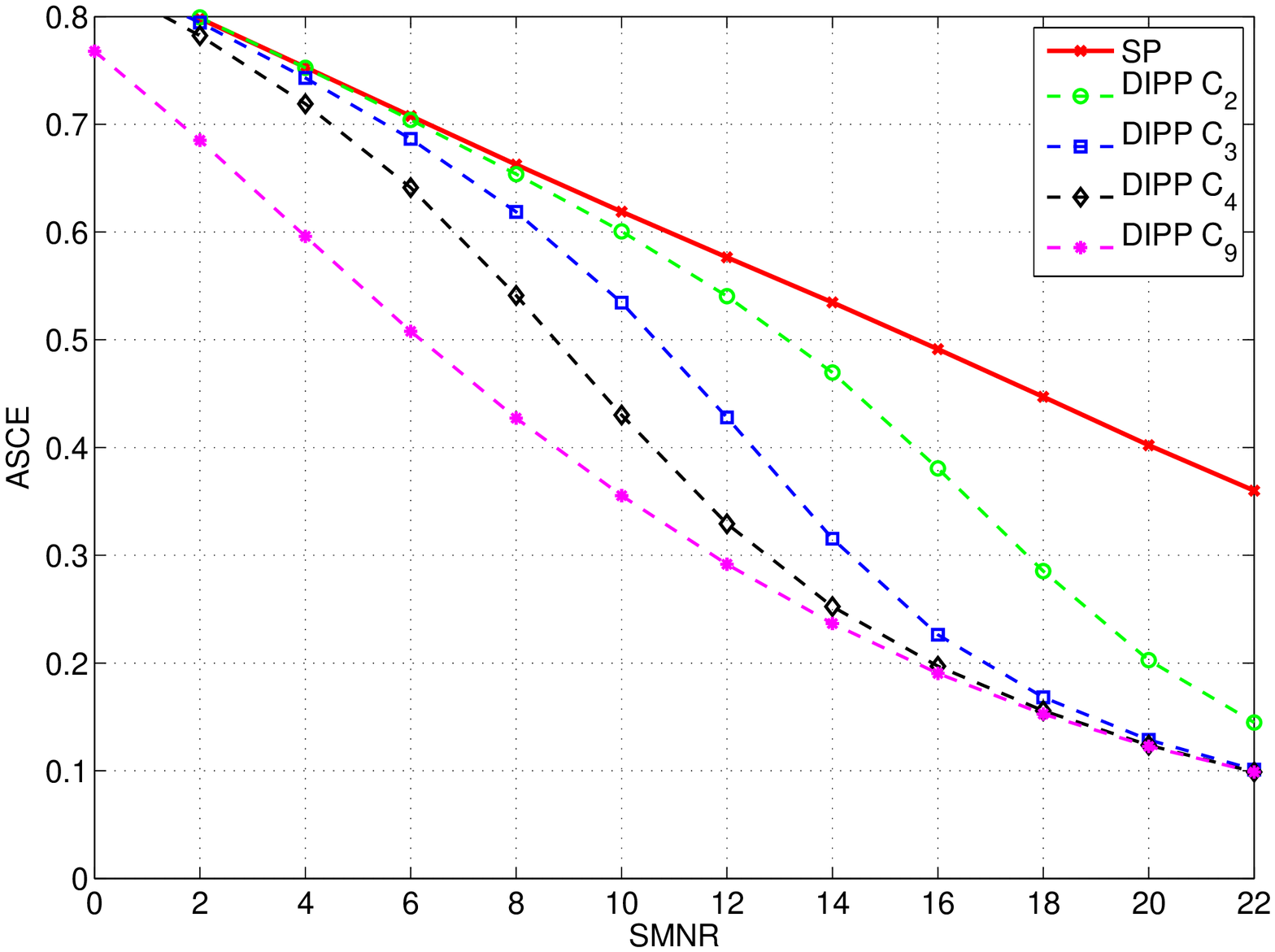} 
      \label{fig:vary_noise_asce}
    }
  }
  \subfloat[\srer vs \smnr]{
    \resizebox{1\columnwidth}{!}{
      \includegraphics[width=\columnwidth]{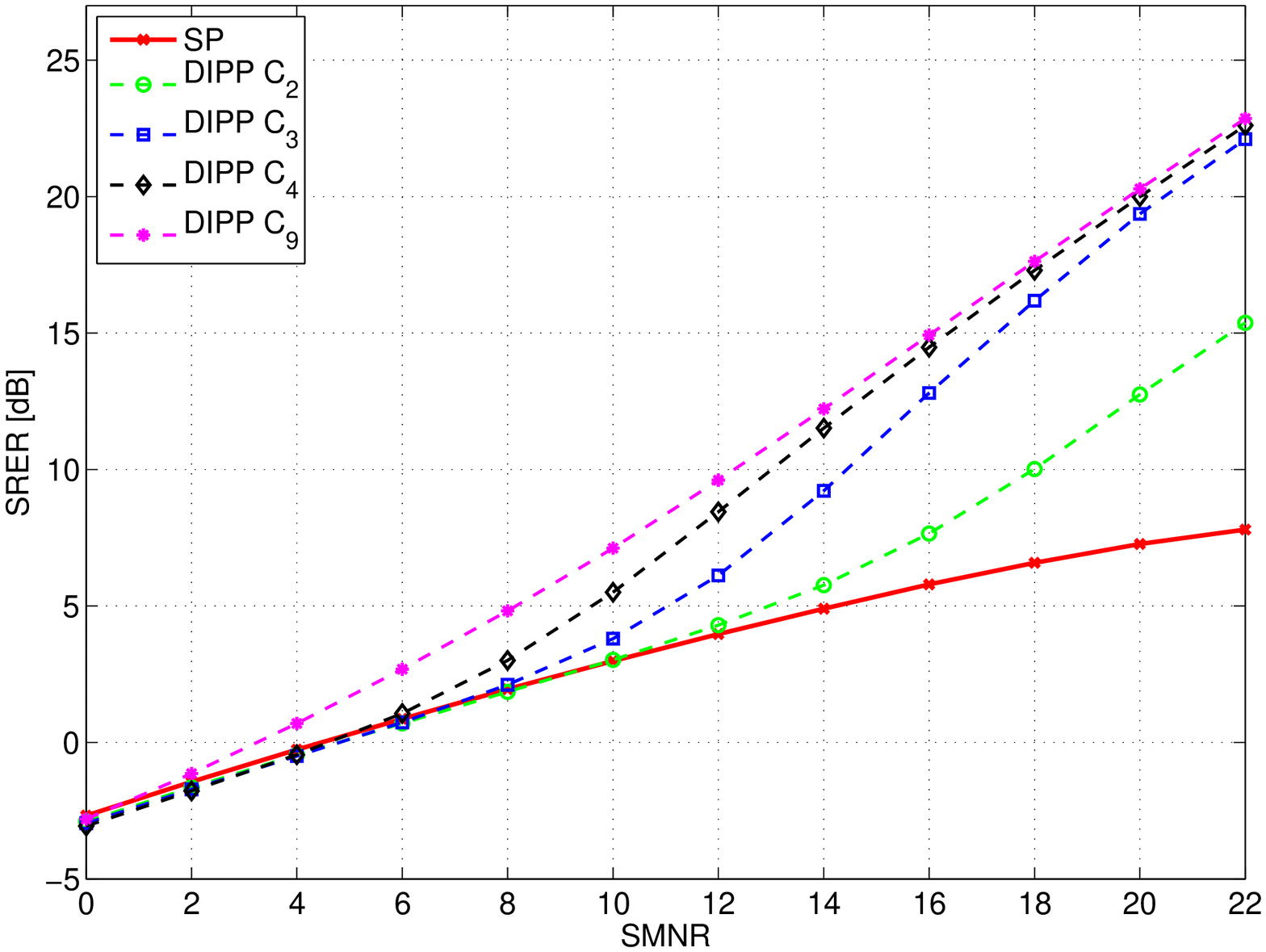}
      \label{fig:vary_noise_srer}
    }
  }
  \caption{{{Reconstruction of Gaussian sparse signal at $\alpha = 0.18$, by varying signal-to-measurement-noise-ratio (\smnr) and network connectivity.}}}
  \label{fig:vary_noise}
\end{figure*}

{{
\subsubsection{Gaussian signals with varying noise}
In \figurename~\ref{fig:vary_noise}, we show results for \srer vs \smnr for various network connectivities with Gaussian sparse signals. In these results, we have chosen $\alpha = 0.18$. Although we notice that better network connectivity consistently provide better results in the higher \smnr region, it is interesting to notice that this is not the case at lower \smnr values. In \figurename~\ref{fig:vary_noise_asce}, we see that a better network connectivity provides better \asce results for all \smnr values. However, in \figurename~\ref{fig:vary_noise_srer}, we see that at the low \smnr region, \sp performs almost the same and sometimes better than \dipp. The likely reason for this interesting result is that estimation of side information may be quite poor in very low \smnr and hence the manadatory inclusion of side information in \dipp may become counter-productive. 
}}

\begin{figure*}[t]
  \centering
  \subfloat[\asce vs fraction of measurements]{
    \resizebox{1\columnwidth}{!}{
      \includegraphics[width=\columnwidth]{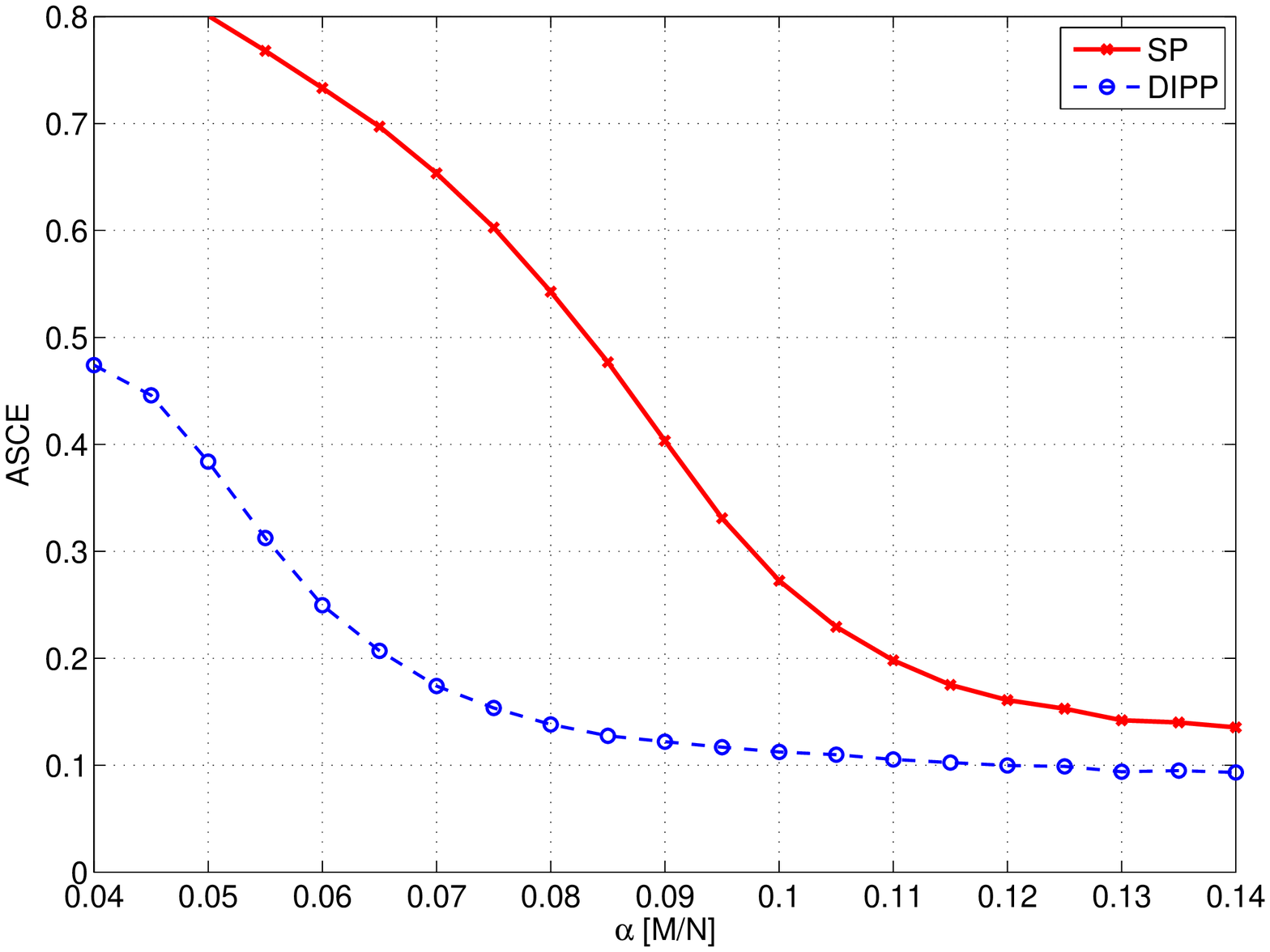} 
      \label{fig:watts_asce}
    }
  }
  \subfloat[\srer vs fraction of measurements]{
    \resizebox{1\columnwidth}{!}{
      \includegraphics[width=\columnwidth]{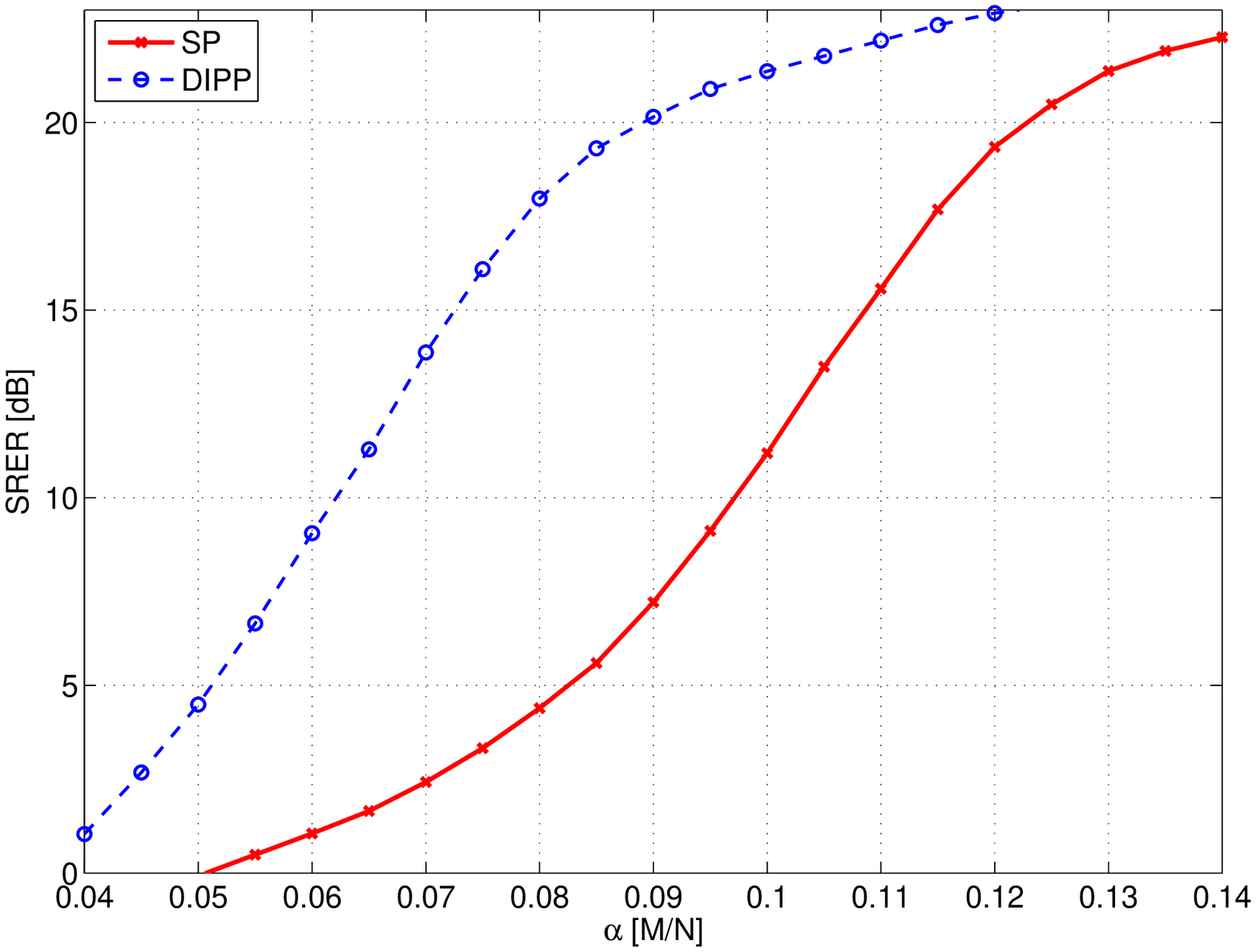}
      \label{fig:watts_srer}
    }
  }
  \caption{{{Reconstruction of Gaussian sparse signal at $\smnr = 20$ dB using the Watts-Strogatz network model~\cite{watts1998collective} with 100 nodes.}}}
  \label{fig:watts}
\end{figure*}

{{
\subsubsection{Large random network}
In \figurename~\ref{fig:watts}, we have simulated the performance for a large random network according to the Watts-Strogatz~\cite{watts1998collective} network model with Gaussian sparse signals at $\smnr = 20$ dB. In this case we have used a total of $100$ nodes, where each node is connected using bidirectional communication links with three other nodes ($q = 3$) according to a circular bi-directional connection strategy. These connections are then with probability $p = 0.3$ rewired to a uniformly chosen random node. We see in \figurename~\ref{fig:watts_asce} and \figurename~\ref{fig:watts_srer} that \dipp provides significantly better performance than \sp.
}}

\textit{Reproducible results:} In the spirit of reproducible results, we provide a package with all necessary \textsc{Matlab} codes in the following website: http://www.ee.kth.se/ctsoftware. In this package consult the \textsc{readme.txt} file to obtain instructions on how to reproduce the figures presented in this paper.

\section{Conclusions}
We show the viability of designing greedy pursuit algorithms for distributed compressed sensing with provable theoretical guarantees based on appropriate assumptions about signal models, sensors and connection networks, as well as providing good practical performance tested via simulations. Through controlled simulations, we show that the developed algorithm follows a natural hypothesis that improvement in network connection leads to improvement in performance. An important conclusion is that a simple voting strategy is efficient to estimate correlation information. 


\appendices
\section{Proof of Lemma~\ref{lemma:bound_complement}} \label{app:lemma:bound_complement}
This lemma is also used in slightly varying forms in other papers, for example in \cite{Ambat:facs,Dai:subspace_pursuit}.

We begin by proving \eqref{lemma:compl1}
  \begin{align}
    \| \mbx - \hat{\mbx} \| & \leq \| \mbx_{\hat{\T}} - \mbA_{\hat{\T}}^{\dagger} \mathbf{y}\| + \| \mbx_{\hat{\T}^{\complement}}\| \nonumber \\ 
    & = \| \mbx_{\hat{\T}} - \mbA_{\hat{\T}}^{\dagger} ( \mbA_{\T}\mbx_{\T} + \mathbf{e} ) \| + \| \mbx_{\hat{\T}^{\complement}}\| \nonumber \\
    & \leq \| \mbx_{\hat{\T}} - \mbA_{\hat{\T}}^{\dagger} \mbA_{\T}\mbx_{\T} \| + \| \mbA_{\hat{\T}}^{\dagger} \mathbf{e} \| + \| \mbx_{\hat{\T}^{\complement}}\| \nonumber \\
    & \overset{(a)}{=} \| \underbrace{\mbx_{\hat{\T}} - \mbA_{\hat{\T}}^{\dagger} \mbA_{\T \cap \hat{\T}} \mbx_{\T \cap \hat{\T}}}_{=0} - \mbA_{\hat{\T}}^{\dagger}\mbA_{\hat{\T}^{\complement}} \mbx_{\hat{\T}^{\complement}} \| \nonumber \\
    & \phantom{= } + \| \mbA_{\hat{\T}}^{\dagger} \mathbf{e} \| + \| \mbx_{\hat{\T}^{\complement}}\| \nonumber \\
    & = \| \mbA_{\hat{\T}}^{\dagger}\mbA_{\hat{\T}^{\complement}} \mbx_{\hat{\T}^{\complement}} \| + \| \mbA_{\hat{\T}}^{\dagger} \mathbf{e} \| + \| \mbx_{\hat{\T}^{\complement}}\| \nonumber \\
    & \overset{(b)}{\leq} \left( 1 + \frac{\delta_{2T}}{1-\delta_{T}} \right) \| \mbx_{\hat{\T}^c} \| + \frac{1}{\sqrt{1-\delta_{T}}} \| \mathbf{e} \| \nonumber\\
    & \overset{(c)}{\leq} \frac{1}{1-\delta_{3T}} \| \mbx_{\hat{\T}^c}
    \| + \frac{1}{\sqrt{1-\delta_{3T}}} \| \mathbf{e} \|.\nonumber
  \end{align}
In the above, we have in \ensuremath{(a)} used that $\mbA_{\T} \mbx_{\T} = \mbA_{\T \cap \hat{\T}} \mbx_{\T \cap \hat{\T}} + \mbA_{\hat{\T}^{\complement}} \mbx_{\hat{\T}^{\complement}}$. In \ensuremath{(b)} we have used proposition~\ref{prop:cosamp}~and~\ref{prop:approx_orthogonality}, and corollary~\ref{cor:approx_orthogonality}. Lastly, in \ensuremath{(c)} we have used $\delta_s \leq \delta_{2T} \leq \delta_{3T}$. Now, proving \eqref{lemma:compl2} is straight forward
\begin{align}
  \| \mbx - \hat{\mbx} \|^2  = \| \mbx_{\hat{\T}} - \mbA_{\hat{\T}}^{\dagger} \mathbf{y}\|^2 + \| \mbx_{\hat{\T}^{\complement}}\|^2 \geq \| \mbx_{\hat{\T}^{\complement}}\|^2 \nonumber
\end{align}
Taking square-root on both sides gives
\begin{align}
  \| \mbx - \hat{\mbx} \| \geq \| \mbx_{\hat{\T}^{\complement}}\|. \nonumber
\end{align}
\qed

\section{Proof of Lemma~\ref{lemma:step1}} \label{app:lemma:step1}
We will prove
\begin{align}
  \| \mbx_{\hat{\T}_l^{\complement}} \| \leq  \frac{1+\delta_{3T}}{1-\delta_{3T}} \|  \mbx_{\check{\mathcal{U}}_l^{\complement}} \| + \frac{2}{\sqrt{1-\delta_{3T}}} \| \mathbf{e} \|. \nonumber
\end{align}
Our proof bears similarity with the proof of Theorem~1 in \cite{Ambat:facs}, \appendixname~I in \cite{Dai:subspace_pursuit} and also \appendixname~C in \cite{Giryes:rip_based_near_oracle_trans_arxiv}. Here, we would like to point out that by following ideas from \cite{Giryes:rip_based_near_oracle_trans_arxiv}, a slightly tighter bound can be formed. However, we abandon those ideas in favor of clarity in the derivations. In this proof we drop the sub-index `$l$' for less notational clutter. We start with defining
  \begin{align}
    \T_{\Delta} = \check{\mathcal{U}} \setminus \hat{\T}. \nonumber
  \end{align}
  Observe that $\hat{\T} \subset \check{\mathcal{U}}$. Then, by using $\hat{\T}^{\complement} = \check{\mathcal{U}}^{\complement} \cup \T_{\Delta}$ we
  get
  \begin{align}
    \| \mbx_{\hat{\T}^{\complement}} \| \leq \|
    \mbx_{\check{\mathcal{U}}^{\complement}} \| + \|
    \mbx_{\T_{\Delta}} \|. \label{eqn:first_ineq}
  \end{align}
  Let us consider the following relation,
    \begin{align}
      \| \check{\mbx}_{\T_{\Delta}} \| & = \| \mbx_{\T_{\Delta}} + (\check{\mbx}_{\check{\mathcal{U}}} - \mbx_{\check{\mathcal{U}}})_{\T_{\Delta}} \| \nonumber \\
      & \geq \| \mbx_{\T_{\Delta}} \| - \|(\check{\mbx}_{\check{\mathcal{U}}} - \mbx_{\check{\mathcal{U}}})_{\T_{\Delta}} \|, \nonumber
    \end{align}
  which by rearranging the terms
    \begin{align}
      \| \mbx_{\T_{\Delta}} \| & \leq \| \check{\mbx}_{\T_{\Delta}} \|  + \| (\check{\mbx}_{\check{\mathcal{U}}} - \mbx_{\check{\mathcal{U}}})_{\T_{\Delta}} \| \nonumber\\
      & \leq \| \check{\mbx}_{\T_{\Delta}} \| + \|\check{\mbx}_{\check{\mathcal{U}}} - \mbx_{\check{\mathcal{U}}}\|. \label{eqn:x_delta}
    \end{align}
  Furthermore, we have that
    \begin{align}
      \| \check{\mbx}_{\T_\Delta} \|^2 & =   \| \check{\mbx}_{\T_\Delta} \|^2 +  \| \check{\mbx}_{\hat{\T}} \|^2 - \| \check{\mbx}_{\hat{\T}} \|^2 \nonumber \\
      & = \| \check{\mbx} \|^2 - \| \check{\mbx}_{\hat{\T}} \|^2 \nonumber \\
      & = \| \check{\mbx}_{\check{\mathcal{U}}\setminus \T} \|^2 + \| \check{\mbx}_{\T} \|^2- \| \check{\mbx}_{\hat{\T}} \|^2 \nonumber \\
      & \overset{(a)}{\leq} \|\check{\mbx}_{\check{\mathcal{U}}\setminus \T} \|^2, \nonumber 
    \end{align}
  where we in \ensuremath{(a)} have used that $\| \check{\mbx}_{\T}
  \|^2- \| \check{\mbx}_{\hat{\T}} \|^2 \leq 0 \text{, by
    definition}$. Taking square-root on both sides gives
    \begin{align}
      \| \check{\mbx}_{\T_\Delta} \| & \leq \| \check{\mbx}_{\check{\mathcal{U}}\setminus \T} \| \nonumber \\
      & = \| \check{\mbx}_{\check{\mathcal{U}}\setminus \T} - \mbx_{\check{\mathcal{U}}\setminus \T} \| \nonumber \\
      & = \| (\check{\mbx}_{\check{\mathcal{U}}} - \mbx_{\check{\mathcal{U}}})_{\check{\mathcal{U}}\setminus \T} \| \nonumber \\
      & \leq \| \check{\mbx}_{\check{\mathcal{U}}} -  \mbx_{\check{\mathcal{U}}} \|. \label{eqn:x_check_delta}
    \end{align}
  Combining \eqref{eqn:x_check_delta} with \eqref{eqn:x_delta} gives
  \begin{align}
    \| \mbx_{\T_{\Delta}} \| \leq 2 \|
    \check{\mbx}_{\check{\mathcal{U}}} - \mbx_{\check{\mathcal{U}}}
    \|. \label{eqn:2_x_minus}
  \end{align}
  Studying RHS of \eqref{eqn:2_x_minus}
    \begin{align}
      \| \mbx_{\check{\mathcal{U}}} - \check{\mbx}_{\check{\mathcal{U}}}\| & = \| \mbx_{\check{\mathcal{U}}} - \mbA_{\check{\mathcal{U}}}^{\dagger}(\mbA\mbx + \mathbf{e}) \| \nonumber \\
      & = \| \mbx_{\check{\mathcal{U}}} - \mbA_{\check{\mathcal{U}}}^{\dagger}(\mbA_{\check{\mathcal{U}}}\mbx_{\check{\mathcal{U}}} + \mbA_{\check{\mathcal{U}}^{\complement}}\mbx_{\check{\mathcal{U}}^{\complement}} + \mathbf{e}) \| \nonumber \\
      & = \| \mbA_{\check{\mathcal{U}}}^{\dagger} \mbA_{\check{\mathcal{U}}^{\complement}}\mbx_{\check{\mathcal{U}}^{\complement}} + \mbA_{\check{\mathcal{U}}}^{\dagger} \mathbf{e} \| \nonumber \\
      & \leq \| \mbA_{\check{\mathcal{U}}}^{\dagger} \mbA_{\check{\mathcal{U}}^{\complement}}\mbx_{\check{\mathcal{U}}^{\complement}}\|  + \| \mbA_{\check{\mathcal{U}}}^{\dagger} \mathbf{e} \| \nonumber \\
      & = \| (\mbA_{\check{\mathcal{U}}}^*\mbA_{\check{\mathcal{U}}})^{-1} \mbA_{\check{\mathcal{U}}}^* \mbA_{\check{\mathcal{U}}^{\complement}}\mbx_{\check{\mathcal{U}}^{\complement}}\|  + \| \mbA_{\check{\mathcal{U}}}^{\dagger} \mathbf{e} \| \nonumber \\
      & \overset{(a)}{\leq} \frac{\delta_{3T}}{1 - \delta_{2T}} \| \mbx_{\check{\mathcal{U}}^{\complement}}\| + \frac{1}{\sqrt{1 - \delta_{2T}}} \| \mathbf{e} \|, \label{eqn:x_minus}
    \end{align}
  where we in \ensuremath{(a)} have used \eqref{prop3.1:5} and
  \eqref{prop3.1:3} of proposition~\ref{prop:cosamp}, and
  corollary~\ref{cor:approx_orthogonality}. Now, by combining
  \eqref{eqn:x_minus}, \eqref{eqn:2_x_minus} and
  \eqref{eqn:first_ineq} we get
  \begin{align}
    \| \mbx_{\hat{\T}^{\complement}} \| & \leq  \left( 1 + \frac{2 \delta_{3T}}{1 - \delta_{2T}} \right) \| \mbx_{\check{\mathcal{U}}^{\complement}}\| + \frac{2}{\sqrt{1- \delta_{2T}}} \| \mathbf{e} \| \nonumber \\
    & \overset{(a)}{\leq} \frac{1+\delta_{3T}}{1-\delta_{3T}} \|
    \mbx_{\check{\mathcal{U}}^{\complement}} \| + \frac{2}{\sqrt{1-
        \delta_{3T}}} \| \mathbf{e} \|, \nonumber
  \end{align}
  where we in \ensuremath{(a)} have used that $\delta_{2T} \leq
  \delta_{3T}$.
\qed

\section{Proof of Lemma~\ref{lemma:sp1}} \label{app:lemma:sp1}
We will show that
\begin{align}
  \| \mbx_{\tilde{\mathcal{U}}_l^{\complement}} \| \leq \frac{2\delta_{3T}}{(1-\delta_{3T})^2} \| \mbx_{\Th_{l-1}^{\complement}} \| + \frac{2\sqrt{1+\delta_{3T}}}{1-\delta_{3T}} \| \mathbf{e} \|. \nonumber
\end{align}
Our proof borrows ideas from proof in \appendixname~H of \cite{Dai:subspace_pursuit} and the proof in \appendixname~B of \cite{Giryes:rip_based_near_oracle_trans_arxiv}. 
We start with  
    \begin{align}
      \mathbf{r}_{l-1} & = \resid(\mby, \mbA_{\Th_{l-1}}) \nonumber \\
      & \overset{(a)}{=} \resid(\mbA_{\Th_{l-1}^{\complement}}\mbx_{\Th_{l-1}^{\complement}}, \mbA_{\Th_{l-1}}) \nonumber \\
      & \phantom{=} + \resid(\mbA_{\Th_{l-1}}\mbx_{\Th_{l-1}}, \mbA_{\Th_{l-1}}) + \resid(\mbe, \mbA_{\Th_{l-1}}) \nonumber \\
      & \overset{(b)}{=} \resid(\mbA_{\Th_{l-1}^{\complement}}\mbx_{\Th_{l-1}^{\complement}}, \mbA_{\Th_{l-1}}) + \resid(\mbe, \mbA_{\Th_{l-1}}). \label{eqn:lemma5_part1}
    \end{align}
    Here, we have in \ensuremath{(a)} used that $\Omega = \Th \cup \Th^{\complement}$ and that the residual operator is linear. In \ensuremath{(b)} we have used that $\resid(\mbA_{\Th_{l-1}}\mbx_{\Th_{l-1}}, \mbA_{\Th_{l-1}}) = \mbA_{\Th_{l-1}}\mbx_{\Th_{l-1}} - \mbA_{\Th_{l-1}}\mbA_{\Th_{l-1}}^{\dagger}\mbA_{\Th_{l-1}}\mbx_{\Th_{l-1}} = 0$. We now study the first term of RHS in \eqref{eqn:lemma5_part1}
    \begin{align}
      & \resid(\mbA_{\Th_{l-1}^{\complement}}\mbx_{\Th_{l-1}^{\complement}}, \mbA_{\Th_{l-1}}) \nonumber \\
      & \phantom{=} = \mbA_{\Th_{l-1}^{\complement}}\mbx_{\Th_{l-1}^{\complement}} - \mbA_{\Th_{l-1}}\mbA_{\Th_{l-1}}^{\dagger}\mbA_{\Th_{l-1}^{\complement}}\mbx_{\Th_{l-1}^{\complement}} \nonumber \\
      & \phantom{=} = \left[\mbA_{\Th_{l-1}^{\complement}} ~  \mbA_{\Th_{l-1}} \right] \left[ \begin{array}{l} \mbx_{\Th_{l-1}^{\complement}} \\ - \mbA_{\Th_{l-1}}^{\dagger}\mbA_{\Th_{l-1}^{\complement}}\mbx_{\Th_{l-1}^{\complement}}\end{array} \right] \nonumber \\
      & \phantom{=} = \mbA \mathbf{q}_{l-1}. \label{eqn:lemma5_residual_nonoise}
    \end{align}
    Observe that $\mathbf{q}_{l-1}$ is at most $2T$-sparse with support-set $\T \cup \Th_{l-1}$. Thus, we can write that
    \begin{align}
      \mbr_{l-1} = \mbA \mathbf{q}_{l-1} + \resid(\mbe, \mbA_{\Th_{l-1}}). \label{eqn:lemma5_residual}
    \end{align}
    Now, by the definition of $\grave{\T}_l$, we have that
    \begin{align}
      & \| \mbA_{\grave{\T}_l}^*\mbr_{l-1} \| \geq \| \mbA_{\T}^*\mbr_{l-1} \| \geq \| \mbA_{\T\setminus \Th_{l-1}}^*\mbr_{l-1} \| \nonumber \\
      & \phantom{=} \overset{(a)}{\geq} \| \mbA_{\T\setminus \Th_{l-1}}^* \mbA \mbq_{l-1}\| - \| \mbA_{\T\setminus \Th_{l-1}}^*\resid(\mbe, \mbA_{\Th_{l-1}}) \| \nonumber \\
      & \phantom{=} \overset{(b)}{\geq} \| \mbA_{\T\setminus \Th_{l-1}}^* \mbA \mbq_{l-1}\| - \sqrt{1 + \delta_T}\|\mbe \|. \label{eqn:lemma3_lowerbound}
    \end{align}
    We have in \ensuremath{(a)} used \eqref{eqn:lemma5_residual} and the reversed triangle inequality, while we in \ensuremath{(b)} have used \eqref{prop3.1:1} in \propositionname~\ref{prop:cosamp} and that $\|\resid(\mbe, \mbA_{\Th_{l-1}})\| \leq \|\mbe\|$. Similarly we can provide the following upper-bound
    \begin{align}
      \| \mbA_{\grave{\T}_l}^*\mbr_{l-1} \| & \overset{(a)}{\leq} \| \mbA_{\grave{\T}_l}^*\mbA \mbq_{l-1} \| + \| \mbA_{\grave{\T}_l}^*\resid(\mbe, \mbA_{\Th_{l-1}}) \| \nonumber \\
      & \overset{(b)}{\leq} \| \mbA_{\grave{\T}_l}^*\mbA \mbq_{l-1} \| + \sqrt{1+\delta_T} \| \mbe \|. \label{eqn:lemma3_upperbound}
    \end{align}
    We have in \ensuremath{(a)} used \eqref{eqn:lemma5_residual} and the triangle inequality, while we in \ensuremath{(b)} have used \eqref{prop3.1:1} in \propositionname~\ref{prop:cosamp} and that $\|\resid(\mbe, \mbA_{\Th_{l-1}})\| \leq \|\mbe\|$.

    Combining \eqref{eqn:lemma3_lowerbound} and \eqref{eqn:lemma3_upperbound} gives
    \begin{align}
      \| \mbA_{\grave{\T}_l}^*\mbA \mbq_{l-1} \| + 2\sqrt{1+\delta_T} \| \mbe \| \geq  \| \mbA_{\T\setminus \Th_{l-1}}^* \mbA \mbq_{l-1}\|. \nonumber
    \end{align}
    We remove the common columns in $\mbA_{\grave{\T}_l}$ and $\mbA_{\T\setminus \Th_{l-1}}$ on both sides (observe that $\Th_{l-1} \not\subseteq \grave{\T}_l$)
    \begin{align}
      \| \mbA_{\grave{\T}_l\setminus \T}^*\mbA \mbq_{l-1} \| + 2\sqrt{1+\delta_T} \| \mbe \| & \geq  \| \mbA_{(\T\setminus \Th_{l-1}) \setminus \grave{\T}_{l}}^* \mbA \mbq_{l-1}\| \nonumber \\
      & \overset{(a)}{=} \| \mbA_{\T\setminus \tilde{\mathcal{U}}_{l}}^* \mbA \mbq_{l-1}\|, \label{eqn:lemma5_ineq_inter_step}
    \end{align}
    where \ensuremath{(a)} follows from that $(\T\setminus \Th_{l-1}) \setminus \grave{\T}_{l} = \T \setminus (\Th_{l-1} \cup \grave{\T}_l) = \T \setminus \tilde{\mathcal{U}}_{l}$. We now upper-bound the first term of the LHS of \eqref{eqn:lemma5_ineq_inter_step}.
    \begin{align}
      \| \mbA_{\grave{\T}_l\setminus \T}^*\mbA \mbq_{l-1} \| & \overset{(a)}{=} \| \mbA_{\grave{\T}_l\setminus \T}^*\mbA_{\T \cup \Th_{l-1}} (\mbq_{l-1})_{\T \cup \Th_{l-1}} \| \nonumber \\
      & \leq \| \mbA_{\grave{\T}_l\setminus \T}^*\mbA_{\T \cup \Th_{l-1}} \| \| (\mbq_{l-1})_{\T \cup \Th_{l-1}} \| \nonumber \\
      & \overset{(b)}{\leq} \delta_{3T} \| \mbq_{l-1} \|. \label{eqn:lemma5_upperbound}
    \end{align}
    In \ensuremath{(a)}, we used that $\mbq_{l-1}$ has support only over $\T \cup \Th_{l-1}$ from \eqref{eqn:lemma5_residual_nonoise}, and in \ensuremath{(b)} we used \propositionname~\ref{prop:approx_orthogonality}. Furthermore, we lower-bound the RHS of \eqref{eqn:lemma5_ineq_inter_step}
    \begin{align}
      & \| \mbA_{\T\setminus \tilde{\mathcal{U}}_{l}}^* \mbA \mbq_{l-1}\| \nonumber \\
      & \overset{(a)}{=} \| \mbA_{\T\setminus \tilde{\mathcal{U}}_{l}}^* \mbA_{\T \cup \Th_{l-1}} (\mbq_{l-1})_{\T \cup \Th_{l-1}}\| \nonumber \\
      & = \| \mbA_{\T\setminus \tilde{\mathcal{U}}_{l}}^* \mbA_{(\T \cup \Th_{l-1})\setminus \tilde{\mathcal{U}}_l} (\mbq_{l-1})_{(\T \cup \Th_{l-1})\setminus \tilde{\mathcal{U}}_l} \nonumber \\
      & \phantom{=} \phantom{=} + \mbA_{\T\setminus \tilde{\mathcal{U}}_{l}}^* \mbA_{(\T \cup \Th_{l-1})\cap \tilde{\mathcal{U}}_l} (\mbq_{l-1})_{(\T \cup \Th_{l-1})\cap \tilde{\mathcal{U}}_l}) \| \nonumber \\
      & \overset{(b)}{\geq} \| \mbA_{\T\setminus \tilde{\mathcal{U}}_{l}}^* \mbA_{\T \setminus \tilde{\mathcal{U}}_l} (\mbq_{l-1})_{\T \setminus \tilde{\mathcal{U}}_l} \| \nonumber \\
      & \phantom{=} \phantom{=} - \| \mbA_{\T\setminus \tilde{\mathcal{U}}_{l}}^* \mbA_{(\T \cup \Th_{l-1})\cap \tilde{\mathcal{U}}_l} (\mbq_{l-1})_{(\T \cup \Th_{l-1})\cap \tilde{\mathcal{U}}_l}) \| \nonumber \\
      & \overset{(c)}{\geq} (1 - \delta_T) \| (\mbq_{l-1})_{\T \setminus \tilde{\mathcal{U}}_l} \| \nonumber \\
      & \phantom{=} \phantom{=} - \| \mbA_{\T\setminus \tilde{\mathcal{U}}_{l}}^* \mbA_{(\T \cup \Th_{l-1})\cap \tilde{\mathcal{U}}_l} \| \| (\mbq_{l-1})_{(\T \cup \Th_{l-1})\cap \tilde{\mathcal{U}}_l}) \| \nonumber \\
      & \overset{(d)}{\geq} (1 - \delta_{3T}) \| (\mbq_{l-1})_{\T \setminus \tilde{\mathcal{U}}_l} \|  - \delta_{3T} \| (\mbq_{l-1})_{(\T \cup \Th_{l-1})\cap \tilde{\mathcal{U}}_l}) \| \nonumber \\
      & \geq (1-\delta_{3T})\| (\mbq_{l-1})_{\T\setminus \tilde{\mathcal{U}}_{l}} \| - \delta_{3T} \| \mbq_{l-1} \|. \label{eqn:lemma5_lowerbound}
\end{align}
In \ensuremath{(a)}, we used that $\mbq_{l-1}$ has support over $\T \cup \Th_{l-1}$. In \ensuremath{(b)}, we used the reversed triangle inequality. In \ensuremath{(c)}, for the first term \eqref{prop3.1:4} of \propositionname~\ref{prop:cosamp} is used and the second term follows from that the spectral norm is sub-multiplicative. In \ensuremath{(d)} \propositionname~\ref{prop:approx_orthogonality} is applied (note that $\T\setminus \tilde{\mathcal{U}}_{l}$ does not intersect with $(\T \cup \Th_{l-1})\cap \tilde{\mathcal{U}}_l$) together with $\delta_T \leq \delta_{2T} \leq \delta_{3T}$.


Now, by substituting \eqref{eqn:lemma5_upperbound} and \eqref{eqn:lemma5_lowerbound} into \eqref{eqn:lemma5_ineq_inter_step}, we get
\begin{align}
  2 \delta_{3T} \| \mbq_{l-1} \| + 2\sqrt{1+\delta_{3T}}\|\mbe \| \geq (1-\delta_{3T}) \| (\mbq_{l-1})_{\T \setminus \tilde{\mathcal{U}}_{l}} \|, \label{eqn:lemma5_last_ineq}
\end{align}
where we also applied that $\delta_{T} \leq \delta_{3T}$. We now observe that
\begin{align}
  \| \mbq_{l-1} \| & = \left\| \left[ \begin{array}{l} \mbx_{\Th_{l-1}^{\complement}} \\ - \mbA_{\Th_{l-1}}^{\dagger}\mbA_{\Th_{l-1}^{\complement}}\mbx_{\Th_{l-1}^{\complement}} \end{array} \right] \right\| \nonumber \\
  & \leq \| \mbx_{\Th_{l-1}^{\complement}} \| + \| \mbA_{\Th_{l-1}}^{\dagger}\mbA_{\Th_{l-1}^{\complement}}\mbx_{\Th_{l-1}^{\complement}} \| \nonumber \\
  & = \| \mbx_{\Th_{l-1}^{\complement}} \| + \| (\mbA_{\Th_{l-1}}^*\mbA_{\Th_{l-1}})^{-1}\mbA_{\Th_{l-1}}^* \mbA_{\Th_{l-1}^{\complement}}\mbx_{\Th_{l-1}^{\complement}} \| \nonumber \\
  & \overset{(a)}{\leq} \| \mbx_{\Th_{l-1}^{\complement}} \| + \frac{\delta_{2T}}{1-\delta_{T}} \| \mbx_{\Th_{l-1}^{\complement}} \| \nonumber \\
  & \overset{(b)}{\leq} \left( 1 + \frac{\delta_{3T}}{1-\delta_{3T}} \right) \| \mbx_{\Th_{l-1}^{\complement}} \| = \frac{1}{1-\delta_{3T}} \| \mbx_{\Th_{l-1}^{\complement}} \|.\label{eqn:lemma5_qpart1}
\end{align}
In \ensuremath{(a)} we have used \eqref{prop3.1:5} of \propositionname~\ref{prop:cosamp} and \corollaryname~\ref{cor:approx_orthogonality}. In \ensuremath{(b)} we have used that $\delta_{T} \leq \delta_{2T} \leq \delta_{3T}$. Furthermore, we have
\begin{align}
  \| (\mbq_{l-1})_{\T \setminus \tilde{\mathcal{U}}_{l}} \| & = \left \| \left[ \begin{array}{l} \mbx_{\Th_{l-1}^{\complement}} \\ - \mbA_{\Th_{l-1}}^{\dagger}\mbA_{\Th_{l-1}^{\complement}}\mbx_{\Th_{l-1}^{\complement}} \end{array} \right]_{\T \setminus \tilde{\mathcal{U}}_l} \right \| \nonumber \\
  & \overset{(a)}{=} \| (\mbx_{\Th_{l-1}^{\complement}})_{\T \setminus \tilde{\mathcal{U}}_l} \| \nonumber \\
  & = \| \mbx_{\tilde{\mathcal{U}}_l^{\complement}} \|, \label{eqn:lemma5_qpart2}
\end{align}
where we in \ensuremath{(a)} have used that $\T \setminus \tilde{\mathcal{U}}_l \subseteq \T \setminus \Th_{l-1} \subset \Th_{l-1}^{\complement}$. By substituting \eqref{eqn:lemma5_qpart1} and \eqref{eqn:lemma5_qpart2} into \eqref{eqn:lemma5_last_ineq}, we get
\begin{align}
  \frac{2\delta_{3T}}{1-\delta_{3T}} \| \mbx_{\Th_{l-1}^{\complement}} \| + 2\sqrt{1+\delta_{3T}}\| \mbe \| \geq (1-\delta_{3T}) \| \mbx_{\tilde{\mathcal{U}}_{l}^{\complement}} \|, \nonumber 
\end{align}
which equivalently can be written as
\begin{align}
  \| \mbx_{\tilde{\mathcal{U}}_{l}^{\complement}} \| \leq \frac{2\delta_{3T}}{(1-\delta_{3T})^2} \| \mbx_{\Th_{l-1}^{\complement}} \| + \frac{2\sqrt{1+\delta_{3T}}}{1-\delta_{3T}}\| \mbe \|. \nonumber 
\end{align}
\qed

\section{Proofs for \dipp} \label{app:dipp}
\begin{IEEEproof}[Proof of Proposition~\ref{prop:recurrence_dipp}]
To find the recurrence inequality for \dipp, we use the performance bound for \sipp and the \expansion jointly, presented in proposition~\ref{thm:perf_bound_pp} and proposition~\ref{thm:merger}. These two propositions have no outer loop iteration parameter $k$. Thus, we introduce proposition~\ref{thm:perf_bound_pp} with the outer loop iteration counter:
\begin{align}
\| (\mbx_{\p})_{\hat{\T}_{\p,k}^{\complement}} \| \leq \frac{b_{\sipp}}{1-a_{\sipp}} \| (\mbx_{\p})_{\T_{\p,\text{si},k}^{\complement}}\| + \frac{1 - a_{\sipp} + c_{\sipp}}{1-a_{\sipp}}  \| \mathbf{e}_{\p} \|. \label{proof:dipp1}
\end{align}
Observe that this relation requires that $a_{\sipp} < 1$. We also introduce proposition~\ref{thm:merger} with iteration counter:
\begin{align}
  \| (\mbx_{\p})_{\T_{\p,\text{si},k}^{\complement}} \| \leq {a_{\texttt{co}}} \| (\mbx_{\p})_{\hat{\T}_{\p,k-1}^{\complement}} \|. \label{proof:dipp2}
\end{align}
By combining \eqref{proof:dipp1} and \eqref{proof:dipp2}, we get 
\begin{align}
  \| (\mbx_{\p})_{\hat{\T}_{\p,k}^{\complement}} \| < {a_{\texttt{co}}} \frac{b_{\sipp}}{1-a_{\sipp}} \| (\mbx_{\p})_{\hat{\T}_{\p,k-1}^{\complement}} \| + \frac{1 - a_{\sipp} + c_{\sipp}}{1-a_{\sipp}}  \| \mathbf{e}_{\p} \|, \nonumber 
\end{align}
where we have assumed that $a_{\sipp} < 1$.
\end{IEEEproof}

{{
\begin{IEEEproof}[Proof of Proposition~\ref{prop:performance_dipp}]
We start with proving~\eqref{dipp:perf1}, by iteratively applying Proposition~\ref{prop:recurrence_dipp}
  \begin{align}
    & \| (\mbx_{\p})_{\hat{\T}_{\p,k}^{\complement}} \| \nonumber \\
    & \phantom{=} \leq  \frac{{a_{\texttt{co}}}b_{\sipp}}{1-a_{\sipp}} \| (\mbx_{\p})_{\hat{\T}_{\p,k-1}^{\complement}} \| + \frac{1 - a_{\sipp} + c_{\sipp}}{1-a_{\sipp}}  \| \mathbf{e}_{\p} \| \nonumber \\
    & \phantom{=} \leq  \frac{{a_{\texttt{co}}} b_{\sipp}}{1-a_{\sipp}} \left( \frac{ {a_{\texttt{co}}} b_{\sipp}}{1-a_{\sipp}} \| (\mbx_{\p})_{\hat{\T}_{\p,k-2}^{\complement}} \| \right. \nonumber \\
    & \phantom{=} \phantom{=} \left. + \frac{1 - a_{\sipp} + c_{\sipp}}{1-a_{\sipp}}  \| \mathbf{e}_{\p} \| \right) + \frac{1 - a_{\sipp} + c_{\sipp}}{1-a_{\sipp}}  \| \mathbf{e}_{\p} \| \nonumber \\
    & \phantom{=} \overset{(a)}{\leq} \left( \frac{{a_{\texttt{co}}}b_{\sipp}}{1-a_{\sipp}} \right)^{k^*} \| (\mbx_{\p})_{\T_{\p,k-k^*}} \| \nonumber \\
    & \phantom{=} \phantom{=} + \frac{1 - a_{\sipp} + c_{\sipp}}{1-a_{\sipp}}
    \sum_{i=0}^{k^*-1} \left(\frac{{a_{\texttt{co}}}b_{\sipp}}{1-a_{\sipp}} \right)^i \|\mathbf{e}_{\p} \|, \nonumber \\
    & \phantom{=} \leq \left( \frac{{a_{\texttt{co}}}b_{\sipp}}{1-a_{\sipp}} \right)^{k^*} \| \mbx_{\p} \| \nonumber \\
    & \phantom{=} \phantom{=} + \frac{1 - a_{\sipp} + c_{\sipp}}{1-a_{\sipp}} 
    \sum_{i=0}^{k^*-1} \left(\frac{{a_{\texttt{co}}}b_{\sipp}}{1-a_{\sipp}} \right)^i \|\mathbf{e}_{\p} \|. \label{app:dipp_p2}
  \end{align}
In \ensuremath{(a)}, we recursively applied the recurrence inequality. We now require that $\frac{{a_{\texttt{co}}}b_{\sipp}}{1-a_{\sipp}} < 1$ such that the first term decays exponentially with iterations $k^*$. Then we plug in $k^*~=~\left\lceil \log\left( \frac{\| \mathbf{e}_{\p} \|}{\|  \mbx_{\p} \|}\right)/\log\left( {a_{\texttt{co}}} \frac{b_{\sipp}}{1-a_{\sipp}} \right) \right\rceil$ into~\eqref{app:dipp_p2} and use geometric series to get that
  \begin{align}
    \| (\mbx_{\p})_{\hat{\T}_{\p,\dipp}^{\complement}} \| & \leq \| \mathbf{e}_{\p} \| + \frac{1 - a_{\sipp} + c_{\sipp}}{1-a_{\sipp}} \frac{1}{1 - \frac{{a_{\texttt{co}}}b_{\sipp}}{1-a_{\sipp}}} \| \mathbf{e}_{\p} \| \nonumber \\
    & = \left( 1 + \frac{1-a_{\sipp} + c_{\sipp}}{1-a_{\sipp}-{a_{\texttt{co}}}b_{\sipp}} \right) \| \mathbf{e}_{\p} \| \nonumber 
  \end{align}
To prove \eqref{dipp:perf2}, we apply lemma~\ref{lemma:bound_complement}
  \begin{align}
    & \| \mbx_{\p} - \hat{\mbx}_{\p,\dipp} \| \nonumber \\
    & \phantom{=} \leq \frac{1}{1-\delta_{3T}} \| (\mbx_{\p})_{\hat{\T}_{\p,\dipp}^c} \| + \frac{1}{\sqrt{1-\delta_{3T}}} \|  \mathbf{e}_{\p} \| \nonumber \\
    & \phantom{=} =  \frac{1}{1-\delta_{3T}} \left( 1 + \frac{1-a_{\sipp} + c_{\sipp}}{1-a_{\sipp}-{a_{\texttt{co}}}b_{\sipp}} \right) \| \mathbf{e}_{\p} \| \nonumber \\
    & \phantom{=} \phantom{=} + \frac{1}{\sqrt{1-\delta_{3T}}} \|  \mathbf{e}_{\p} \| \nonumber \\
    & \phantom{=} \leq \left( \frac{1-a_{\sipp} + c_{\sipp}}{(1-\delta_{3T})(1-a_{\sipp}-{a_{\texttt{co}}}b_{\sipp})} \right. \nonumber \\
    & \phantom{=} \phantom{=} \left. + \frac{2}{1-\delta_{3T}} \right) \|
    \mathbf{e}_{\p} \|, \nonumber
  \end{align}
which concludes the proof.
\end{IEEEproof}
}}

\ifCLASSOPTIONcaptionsoff
  \newpage
\fi

\bibliographystyle{IEEEtran}
\bibliography{references/IEEEfull,references/myconffull,references/compressed_sensing,references/biblio_saikat_CS1,references/biblio_saikat_Pub,references/General,references/references_short}

\begin{thebibliography}{10}
\providecommand{\url}[1]{#1}
\csname url@samestyle\endcsname
\providecommand{\newblock}{\relax}
\providecommand{\bibinfo}[2]{#2}
\providecommand{\BIBentrySTDinterwordspacing}{\spaceskip=0pt\relax}
\providecommand{\BIBentryALTinterwordstretchfactor}{4}
\providecommand{\BIBentryALTinterwordspacing}{\spaceskip=\fontdimen2\font plus
\BIBentryALTinterwordstretchfactor\fontdimen3\font minus
  \fontdimen4\font\relax}
\providecommand{\BIBforeignlanguage}[2]{{%
\expandafter\ifx\csname l@#1\endcsname\relax
\typeout{** WARNING: IEEEtran.bst: No hyphenation pattern has been}%
\typeout{** loaded for the language `#1'. Using the pattern for}%
\typeout{** the default language instead.}%
\else
\language=\csname l@#1\endcsname
\fi
#2}}
\providecommand{\BIBdecl}{\relax}
\BIBdecl

\bibitem{Sundman:parallel_pursuit_for_dcs}
D.~Sundman, S.~Chatterjee, and M.~Skoglund, ``Parallel pursuit for distributed
  compressed sensing,'' in \emph{IEEE Global Conference on Signal and
  Information Processing (GlobalSIP)}, Austin, Texas, USA, Dec. 2013, pp.~--.

\bibitem{Donoho:compressed_sensing}
D.~Donoho, ``Compressed sensing,'' \emph{{IEEE} Transactions on Information
  Theory}, vol.~52, pp. 1289--1306, Apr. 2006.

\bibitem{Candes:stable_signal_recovery}
E.~J. Cand\`{e}s, J.~Romberg, and T.~Tao, ``Stable signal recovery from
  incomplete and inaccurate measurements,'' \emph{Communications on Pure and
  Applied Mathematics}, vol.~59, pp. 1207--1223, Aug. 2006.

\bibitem{Mota:distributed_basis_pursuit}
J.~Mota, J.~Xavier, P.~Aguiar, and M.~Puschel, ``Distributed basis pursuit,''
  \emph{{IEEE} Transactions on Signal Processing}, vol.~60, pp. 1942--1956,
  Apr. 2012.

\bibitem{Bazerque:distributed_spectrum_sensing}
J.~Bazerque and G.~Giannakis, ``Distributed spectrum sensing for cognitive
  radio networks by exploiting sparsity,'' \emph{{IEEE} Transactions on Signal
  Processing}, vol.~58, pp. 1847--1862, Mar. 2010.

\bibitem{Ji:bayesian_compressive_sensing}
S.~Ji, Y.~Xue, and L.~Carin, ``Bayesian compressive sensing,'' \emph{{IEEE}
  Transactions on Signal Processing}, vol.~56, pp. 2346--2356, Jun. 2008.

\bibitem{Donoho:message_passing_for_cs}
D.~Donoho, A.~Maleki, and A.~Montanari, ``Message-passing algorithms for
  compressed sensing,'' \emph{Proceedings of the National Academy of Sciences},
  vol. 106, pp. 18\,914--18\,919, Oct. 2009.

\bibitem{Mallat:matching_pursuit_with_time_frequency_dictionaries}
S.~Mallat and Z.~Zhang, ``Matching pursuits with time-frequency dictionaries,''
  \emph{{IEEE} Transactions on Signal Processing}, vol.~41, pp. 3397--3415,
  Dec. 1993.

\bibitem{Tropp:signal_recovery}
J.~Tropp and A.~Gilbert, ``Signal recovery from random measurements via
  orthogonal matching pursuit,'' \emph{{IEEE} Transactions on Information
  Theory}, vol.~53, pp. 4655--4666, Dec. 2007.

\bibitem{Donoho:sparse_solution_of_underdetermined_systems_stomp}
D.~Donoho, Y.~Tsaig, I.~Drori, and J.-L. Starck, ``Sparse solution of
  underdetermined systems of linear equations by stagewise orthogonal matching
  pursuit,'' \emph{{IEEE} Transactions on Information Theory}, vol.~58, pp.
  1094--1121, Feb. 2012.

\bibitem{Chatterjee:projection_based_look_ahead}
S.~Chatterjee, D.~Sundman, M.~Vehkaper{\"a}, and M.Skoglund, ``Projection-based
  and look-ahead strategies for atom selection,'' \emph{{IEEE} Transactions on
  Signal Processing}, vol.~60, pp. 634--647, Feb. 2012.

\bibitem{Sundman:frogs}
D.~Sundman, S.~Chatterjee, and M.~Skoglund, ``Frogs: A serial reversible greedy
  search algorithm,'' in \emph{IEEE Swedish Communication Technologies Workshop
  (Swe-CTW)}, Lund, Sweden, Mar. 2012, pp. 40--45.

\bibitem{Needell:signal_recovery_from_incomplete_and_inaccurate_measurements_v%
ia_romp}
D.~Needell and R.~Vershynin, ``Signal recovery from incomplete and inaccurate
  measurements via regularized orthogonal matching pursuit,'' \emph{{IEEE}
  Journal of Selected Topics in Signal Processing}, vol.~4, pp. 310--316, Mar.
  2010.

\bibitem{Needell:cosamp}
D.~Needell and J.~A. Tropp, ``Cosamp: Iterative signal recovery from incomplete
  and inaccurate samples,'' \emph{Applied and Computational Harmonic Analysis},
  vol.~26, pp. 301--321, Apr. 2009.

\bibitem{Dai:subspace_pursuit}
W.~Dai and O.~Milenkovic, ``Subspace pursuit for compressive sensing signal
  reconstruction,'' \emph{{IEEE} Transactions on Information Theory}, vol.~55,
  pp. 2230--2249, May 2009.

\bibitem{Sundman:look_ahead_parallel_pursuit}
D.~Sundman, S.~Chatterjee, and M.~Skoglund, ``Look ahead parallel pursuit,'' in
  \emph{IEEE Swedish Communication Technologies Workshop (Swe-CTW)}, Stockholm,
  Sweden, Oct. 2011, pp. 114--117.

\bibitem{Vaswani_KF_CS_2008}
N.~Vaswani, ``Kalman filtered compressed sensing,'' in \emph{Image Processing,
  2008. ICIP 2008. 15th IEEE International Conference on}, oct. 2008, pp. 893
  --896.

\bibitem{Carmi_KF_CS_2010}
A.~Carmi, P.~Gurfil, and D.~Kanevsky, ``Methods for sparse signal recovery
  using kalman filtering with embedded pseudo-measurement norms and
  quasi-norms,'' \emph{Signal Processing, IEEE Transactions on}, vol.~58,
  no.~4, pp. 2405 --2409, april 2010.

\bibitem{Zachariah_Chatterjee_Jansson_TSP_2012}
D.~Zachariah, S.~Chatterjee, and M.~Jansson, ``Dynamic iterative pursuit,''
  \emph{Signal Processing, IEEE Transactions on}, vol.~60, no.~9, pp. 4967
  --4972, sept. 2012.

\bibitem{Yang:distributed_perception}
A.~Yang, M.~Gastpar, R.~Bajcsy, and S.~Sastry, ``Distributed sensor perception
  via sparse representation,'' \emph{Proceedings of the {IEEE}}, vol.~98, pp.
  1077--1088, Jun. 2010.

\bibitem{Feng:distributed_compressive_spectrum_sensing}
F.~Zeng, C.~Li, and Z.~Tian, ``Distributed compressive spectrum sensing in
  cooperative multihop cognitive networks,'' \emph{{IEEE} Transactions on
  Signal Processing}, vol.~5, pp. 37--48, Feb. 2011.

\bibitem{Bazeraque:distributed_spectrum_sensing}
J.~Bazerque and G.~Giannakis, ``Distributed spectrum sensing for cognitive
  radio networks by exploiting sparsity,'' \emph{{IEEE} Transactions on Signal
  Processing}, vol.~58, pp. 1847--1862, Mar. 2010.

\bibitem{Ling:decenteralized_support_detection}
Q.~Ling and T.~Zhi, ``Decentralized support detection of multiple measurement
  vectors with joint sparsity,'' in \emph{IEEE International Conference on
  Acoustics, Speech and Signal Processing (ICASSP)}, Prague, Czech Republic,
  May 2011, pp. 2996--2999.

\bibitem{Sundman:psd}
D.~Sundman, S.~Chatterjee, and M.~Skoglund, ``On the use of compressive
  sampling for wide-band spectrum sensing,'' in \emph{IEEE International
  Symposium on Signal Processing and Information Technology (ISSPIT)}, Luxor,
  Egypt, Dec. 2010, pp. 354--359.

\bibitem{Tropp:simultaneous_sparse_approx_part1}
J.~Tropp, A.~Gilbert, and M.~Strauss, ``Algorithms for simultaneous sparse
  approximation. part i: Greedy pursuit,'' \emph{Signal Processing}, vol.~86,
  pp. 572--588, Mar. 2006.

\bibitem{Rakotomamonjy:surveying}
A.~Rakotomamonjy, ``Surveying and comparing simultaneous sparse approximation
  (or group-lasso) algorithms,'' \emph{Signal Processing}, vol.~91, pp.
  1505--1526, Jul. 2011.

\bibitem{Leviatan:simultaneous}
D.~Leviatan and V.~Temlyakov, ``Simultaneous approximation by greedy
  algorithms,'' \emph{Advances in Computational Mathematics}, vol.~25, pp.
  73--90, Jul. 2006.

\bibitem{Cotter:sparse_solutions}
S.~Cotter, B.~Rao, K.~Engan, and K.~Kreutz-Delgado., ``Sparse solutions to
  linear inverse problems with multiple measurement vectors,'' \emph{{IEEE}
  Transactions on Signal Processing}, vol.~53, pp. 2477--2488, Jul. 2005.

\bibitem{Chen:theoretical_results_on_sparse}
J.~Chen and X.~Huo, ``Theoretical results on sparse representations of
  multiple-measurement vectors,'' \emph{{IEEE} Transactions on Signal
  Processing}, vol.~54, pp. 4634--4643, Dec. 2006.

\bibitem{Sundman:greedy_pursuit_for_jointly}
D.~Sundman, S.~Chatterjee, and M.~Skoglund, ``Greedy pursuits for compressed
  sensing of jointly sparse signals,'' in \emph{EURASIP European Signal
  Processing Conference (EUSIPCO)}, Barcelona, Spain, Aug. 2011, pp. 368--372.

\bibitem{Boyd_ADMM_2011}
S.~Boyd, N.~Parikh, E.~Chu, B.~Peleato, and J.~Eckstein, ``Distributed
  optimization and statistical learning via the alternating method of
  multipliers,'' \emph{Foundations and Trends in Machine Learning}, vol.~3,
  no.~1, pp. 1--122, 2011.

\bibitem{Sundman:diprsp}
D.~Sundman, D.~Zachariah, and S.~Chatterjee, ``Distributed predictive subspace
  pursuit,'' in \emph{IEEE International Conference on Acoustics, Speech and
  Signal Processing (ICASSP)}, Vancouver, Canada, May 2013, pp. 4633--4637.

\bibitem{Sundman:a_greedy_pursuit_algorithm}
D.~Sundman, S.~Chatterjee, and M.~Skoglund, ``A greedy pursuit algorithm for
  distributed compressed sensing,'' in \emph{IEEE International Conference on
  Acoustics, Speech and Signal Processing (ICASSP)}, Kyoto, Japan, Mar. 2012,
  pp. 2729--2732.

\bibitem{Sundman:distributed_gp_algorithms}
------, ``Distributed greedy pursuit algorithms,'' \emph{Signal Processing},
  vol. 105, pp. 298--315, Dec. 2014.

\bibitem{Wimalajeewa:cooperative}
T.~Wimalajeewa and P.~Varshney, ``Cooperative sparsity pattern recovery in
  distributed networks via distributed-omp,'' in \emph{IEEE International
  Conference on Acoustics, Speech and Signal Processing (ICASSP)}, Vancouver,
  Canada, May 2013, pp. 5288--5292.

\bibitem{Zhang:side_information_based_omp}
W.~Zhang, C.~Ma, W.~Wang, Y.~Liu, and L.~Zhang, ``Side information based
  orthogonal matching pursuit in distributed compressed sensing,'' in
  \emph{IEEE International Conference on Network Infrastructure and Digital
  Content}, Beijing, China, Jan. 2010, pp. 80--84.

\bibitem{Elad_book_2010}
M.~Elad, \emph{Sparse and redundant representations: From theory to
  applications in signal and image processing}.\hskip 1em plus 0.5em minus
  0.4em\relax Springer, 2010.

\bibitem{Candes:decoding_linear_programming}
E.~Cand\`{e}s and T.~Tao, ``Decoding by linear programming,'' \emph{{IEEE}
  Transactions on Information Theory}, vol.~51, pp. 4203--4215, Dec. 2005.

\bibitem{Nishimori-2001}
H.~Nishimori, \emph{Statistical Physics of Spin Glasses and Information
  Processing}.\hskip 1em plus 0.5em minus 0.4em\relax New York: Oxford
  University Press, 2001.

\bibitem{Rangan_Replica_CS_TIT_2012}
S.~Rangan, A.~Fletcher, and V.~Goyal, ``Asymptotic analysis of map estimation
  via the replica method and applications to compressed sensing,''
  \emph{Information Theory, IEEE Transactions on}, vol.~58, no.~3, pp. 1902
  --1923, march 2012.

\bibitem{Kabashima_Vehkapera_Chatterjee_2012}
\BIBentryALTinterwordspacing
Y.~Kabashima, M.~Vehkaper\"{a}, and S.~Chatterjee, ``Typical $l_1$-recovery
  limit of sparse vectors represented by concatenations of random orthogonal
  matrices,'' \emph{Journal of Statistical Mechanics: Theory and Experiment},
  vol. 2012, no.~12, p. P12003, 2012. [Online]. Available:
  \url{http://stacks.iop.org/1742-5468/2012/i=12/a=P12003}
\BIBentrySTDinterwordspacing

\bibitem{Vehkapera_Kabashima_Chatterjee_TIT_2014}
\BIBentryALTinterwordspacing
M.~Vehkaper\"{a}, Y.~Kabashima, and S.~Chatterjee, ``Analysis of regularized ls
  reconstruction and random matrix ensembles in compressed sensing,''
  \emph{CoRR}, December 2013. [Online]. Available:
  \url{http://arxiv.org/abs/1312.0256}
\BIBentrySTDinterwordspacing

\bibitem{Davenport_2010_Orthogonal_Matching_pursuit}
M.~Davenport and W.~Wakin, ``Analysis of orthogonal matching pursuit using the
  restricted isometry property,'' \emph{Information Theory, IEEE Transactions
  on}, vol.~56, no.~9, pp. 4395 --4401, sept. 2010.

\bibitem{Baraniuk_Model_Based_CS_TIT_2010}
R.~Baraniuk, V.~Cevher, M.~Duarte, and C.~Hegde, ``Model-based compressive
  sensing,'' \emph{Information Theory, IEEE Transactions on}, vol.~56, no.~4,
  pp. 1982 --2001, april 2010.

\bibitem{Ambat:facs}
S.~Ambat, S.~Chatterjee, and K.~Hari, ``Fusion of algorithms for compressed
  sensing,'' \emph{{IEEE} Transactions on Signal Processing}, vol.~61, pp.
  3699--3704, Apr. 2013.

\bibitem{Ambat_Chatterjee_Hari_2013_CoMACS}
------, ``A committee machine approach for compressed sensing reconstruction,''
  \emph{Signal Processing, IEEE Transactions on}, vol.~62, no.~7, pp.
  1705--1717, 2014.

\bibitem{Kirmani:codac}
A.~Kirmani, A.~Colaco, F.~Wong, and V.~Goyal, ``{C}o{DAC}: A compressive depth
  acquisition camera framework,'' in \emph{IEEE International Conference on
  Acoustics, Speech and Signal Processing (ICASSP)}, Kyoto, Japan, Mar. 2012,
  pp. 5425--5428.

\bibitem{Wu:spherical_microphone}
P.~Wu, N.~Epain, and C.~Jin, ``A dereverberation algorithm for spherical
  microphone arrays using compressed sensing techniques,'' in \emph{IEEE
  International Conference on Acoustics, Speech and Signal Processing
  (ICASSP)}, Kyoto, Japan, Mar. 2012, pp. 4053--4056.

\bibitem{watts1998collective}
D.~J. Watts and S.~H. Strogatz, ``Collective dynamics of 'small-world'
  networks,'' \emph{Nature}, vol. 393, pp. 440--442, 1998.

\bibitem{Giryes:rip_based_near_oracle_trans}
R.~Giryes and M.~Elad, ``Rip-based near-oracle performance guarantees for sp,
  cosamp, and iht,'' \emph{{IEEE} Transactions on Signal Processing}, vol.~60,
  pp. 1465--1468, Mar. 2012.

\bibitem{Sundman:democratic_vote}
\BIBentryALTinterwordspacing
D.~Sundman, S.~Chatterjee, and M.~Skoglund, ``Analysis of democratic voting
  principles used in distributed greedy algorithms,'' \emph{CoRR}, 2014.
  [Online]. Available: \url{http://arxiv.org/abs/1407.4491}
\BIBentrySTDinterwordspacing

\bibitem{Gastpar:a_note_on_optimal_support_recovery}
G.~Reeves and M.~Gastpar, ``A note on optimal support recovery in compressed
  sensing,'' in \emph{Annual Asilomar Conference on Signals, Systems, and
  Computers}, Pacific Grove, USA, Nov. 2009, pp. 1576--1580.

\bibitem{Candes:an_introduction}
E.~Cand\`{e}s and M.~Wakin, ``An introduction to compressive sampling,''
  \emph{{IEEE} Electron Device Letters}, vol.~25, pp. 21--30, Mar. 2008.

\bibitem{Giryes:rip_based_near_oracle_trans_arxiv}
\BIBentryALTinterwordspacing
R.~Giryes and M.~Elad, ``Rip-based near-oracle performance guarantees for
  subspace-pursuit, cosamp, and iterative hard-thresholding,'' \emph{ArXiv
  e-prints}, vol. abs/1005.4539, May 2010. [Online]. Available:
  \url{http://arxiv.org/abs/1005.4539}
\BIBentrySTDinterwordspacing

\end{thebibliography}

\end{document}